\documentclass[12pt]{article}

\usepackage{amsmath, amssymb, amsfonts,lipsum}
\usepackage{hyperref}
\usepackage{graphicx, subfigure, epstopdf}
\usepackage{color}

\begin{document}

\pagenumbering{gobble}
\thispagestyle{empty}

\title{
	{\bf A strong electroweak phase transition\\
	 in the 2HDM after LHC8}\\[3mm]
	\author{ \normalsize {\bf G.~C.~Dorsch}\thanks{G.Dorsch@sussex.ac.uk}~, 
{\bf S.~J.~Huber}\thanks{S.Huber@sussex.ac.uk}~, {\bf J.~M.~No}\thanks{J.M.No@sussex.ac.uk}\\
	\footnotesize \it Department of Physics and Astronomy, University of Sussex, Brighton, BN1 9QH, UK}
}
\date{}
\maketitle

\abstract{The nature of the electroweak phase transition in two-Higgs-doublet models is revisited in 
light of the recent LHC results. A scan over an extensive region of their parameter space is performed, 
showing that a strongly first-order phase transition favours a light neutral scalar with SM-like properties, 
together with a heavy pseudo-scalar ($m_{A^0}\gtrsim 400$~GeV) and a mass hierarchy in the scalar 
sector, $m_{H^\pm}\lesssim m_{H^0} < m_{A^0}$. We also investigate the $h^0\to\gamma\gamma$ decay channel and 
find that an enhancement in the branching ratio is allowed, and in some cases even preferred, when a strongly 
first-order phase transition is required.}

\newpage
\pagenumbering{arabic}
\setcounter{page}{1}
\renewcommand{\theequation}{\arabic{section}.\arabic{equation}}

\section{Introduction}

The baryon asymmetry of the Universe (BAU) is one of the few observables whose experimental value cannot be accounted 
for within the Standard Model (SM). The model could potentially contain all necessary ingredients for the dynamical 
generation of a baryonic excess (namely C, CP and baryon number violation occurring 
out of thermodynamical equilibrium~\cite{Sakharov:1967dj}), since the rate of baryon-number violating processes 
in the SM in unsupressed for temperatures above the electroweak (EW) scale~\cite{Kuzmin:1985mm}. However, 
the departure from thermal equilibrium would have to come from a strong first-order electroweak phase transition, 
and in the SM this would be satisfied only for a Higgs mass $m_h \lesssim m_W$~\cite{Kajantie:1996mn,Csikor:1998eu}. 
Moreover, the amount of CP violation in the SM also turns out to be insufficient for the generation of the 
BAU (for recent reviews, see e.g.~\cite{Cline:2006ts,Morrissey:2012db,Konstandin:2013caa}). 

The BAU is then a strong, empirically-based motivation to look for physics beyond the SM, and focusing on the 
problem of the strength of the electroweak phase transition, it may be natural to extend the scalar sector of the SM. 
The era of direct probe of scalars' properties has just been inaugurated with the recent discovery of a 
spin-$0$ resonance of mass $m_h \sim 125$~GeV at the LHC~\cite{Aad:2012tfa, Chatrchyan:2012ufa}, and even though 
this finding can be rightly celebrated as another success of the SM, the properties of this resonance have not 
yet been probed well enough to establish whether it behaves exactly as the SM scalar boson or not. An extension of 
the SM scalar sector is thus not only allowed, but also testable at present and in the near future, and may even come 
to be desirable, depending on the upcoming results of ATLAS and CMS.

In this paper we study the nature of the electroweak phase transition in two-Higgs-doublet models (2HDMs). These 
are very minimal scalar extensions of the SM, differing from it only by the addition of an extra scalar $SU(2)_L$ 
doublet. Thus, apart from the three Goldstone bosons and the usual Higgs field, the model 
contains a charged and two additional neutral scalar particles, whose interactions with the Higgs result in new 
contributions to its finite temperature effective potential that strengthen the phase transition. Moreover, the 
model can also have additional sources of CP violation, either explicitly in the scalar potential or spontaneously 
via a relative phase between the VEVs of the doublets\footnote{Spontaneous CP violation was the motivation for 
the introduction of 2HDMs by T. D. Lee~\cite{Lee:1974jb}.}. Thus, despite being a very minimal extension of the 
SM, 2HDMs have all that is needed to boost the SM prediction of the BAU.

In fact, previous studies of baryogenesis in 2HDMs confirm that the model can predict the measured value of 
the BAU in some simplified cases~\cite{McLerran:1990zh,Turok:1990zg}, for specific regions of its parameter 
space~\cite{Cohen:1991iu,Cline:1996mga,Fromme:2006cm} and in the most general, 
CP-violating scenario~\cite{Cline:2011mm}. A general study of the dependence of the electroweak phase 
transition with the various parameters of a 2HDM is however challenging, due to the high dimensionality of 
the parameter space of the model --- 14 free parameters in the most general case~\cite{Cline:2011mm}, reduced 
to 10 when a softly broken $\mathbb{Z}_2$ symmetry is imposed. 

The purpose of this work is precisely to shed some light on this issue, which remains so far largely unexplored. 
We perform a random scan over an extensive region of the parameter space looking for points with a strong first-order electroweak phase transition. Our aim is then to establish the extent to which 2HDMs are viable models for baryogenesis, and which regions of the parameter space are preferred for this purpose. Also, in light of the recent experimental data from the LHC, we analyze the interplay between a strongly first-order electroweak phase transition in 2HDMs and several properties of the light neutral scalar $h^0$, such as the deviation of its coupling to weak gauge bosons ($W$ and $Z$) from the SM value and the behaviour of the decay channel $h^0 \rightarrow \gamma \gamma$ in these models. 

Recently there has been considerable activity in linking LHC data with the electroweak phase transition in 
extensions of the SM, for instance in the MSSM~\cite{Carena:2012np} or in singlet 
models~\cite{Fairbairn:2013uta, Damgaard:2013kva}. Typically a tension is found between the requirement 
of a strong first-order phase transition and a SM-like light Higgs. Here, on the contrary, we show 
that a strong first-order phase transition favours a SM-like light Higgs in the framework of 2HDMs. 
 
The paper is organized as follows: In Section~\ref{section:model} we 
review the general features of the 2HDM, including a brief discussion of electroweak 
precision constraints, and describe our method for obtaining the strength of the electroweak phase transition via 
the finite temperature effective potential. Section~\ref{section:scan} describes the details of our scan.  
In Section~\ref{section:data} we discuss the constraints from flavour observables, present our 
analysis of the data sample and confront the results with the latest data from LHC8. Our conclusions will 
be presented in Section~\ref{section:conclusions}.

\section{\label{section:model}The model}

\subsection{General Properties of 2HDMs}

In general the existence of two doublets coupling to the fermions opens an undesirable window 
for FCNC's at tree-level, since the most general fermionic Lagrangian will involve terms of the form
\begin{equation}
\mathcal{L}_{Yukawa}=-\overline{Q}_L\left(\Gamma_1\Phi_1 + \Gamma_2\Phi_2\right)q_R+\ldots
\end{equation}
and diagonalization of the quark mass matrix $M_q = \Gamma_1\langle\Phi_1\rangle + \Gamma_2\langle\Phi_2\rangle$ does 
not imply diagonalization of $\Gamma_1$ and $\Gamma_2$ separately. A very convenient way to deal with this 
problem\footnote{Other ways include Minimal Flavour 
Violation~\cite{Branco:1996bq, D'Ambrosio:2002ex, Buras:2010mh} and general 
Yukawa alignment~\cite{Pich:2009sp}, to name a few.}, which is also
most widely adopted in the literature, is to consider that each type of fermion couples to one 
doublet only~\cite{Glashow:1976nt}. This can be achieved by imposing a $\mathbb{Z}_2$ 
symmetry on the Lagrangian, under which the fields 
transform as $\Phi_1\rightarrow -\Phi_1$, $d_R\rightarrow \pm d_R$ and $l_R\rightarrow \pm l_R$ (the other 
fields remaining invariant). By convention, up-type quarks always 
couple to $\Phi_2$ (so $u_R$ is even under $\mathbb{Z}_2$), but which doublet couples to leptons and 
down-type quarks may vary. There are, accordingly, four such possibilities, and 2HDMs are often categorized, 
according to this choice, as Type I, Type II, Type X and Type Y, as shown in Table~\ref{tab:Types}. 
\begin{table}[h!]
   \centering
   \begin{tabular}{|c|c|c|c|}
       \hline
        \ & $u_R$ & $d_R$ & $e_R$ \\ \hline
	Type I & $+$ & $+$ & $+$ \\ \hline
	Type II & $+$ & $-$ & $-$ \\ \hline
	Type X & $+$ & $+$ & $-$ \\ \hline
	Type Y & $+$ & $-$ & $+$ \\ \hline
  \end{tabular}
  \caption{\small{$\mathbb{Z}_2$ charge of fermions in the different 2HDM Types.}}
   \label{tab:Types}
\end{table}

For the electroweak phase transition the particular type of model is irrelevant, since 
only the top-quark coupling plays a significant role in these effects. Thus, for all our purposes the only 
difference between these models are the relevant experimental constraints, resulting in different allowed 
regions of their parameter space (cf. Section~\ref{section:pheno_constraints}).  

The most general gauge-invariant and renormalizable potential that can be written for two doublets under 
the previous conditions, allowing for a soft breaking of the $\mathbb{Z}_2$ symmetry, is
\begin{equation}	
	\label{2HDM_potential}
	\begin{split}
		V_{\rm tree}(\Phi_1,\Phi_2)=&-\mu^2_1 \Phi_1^{\dagger}\Phi_1-
					  \mu^2_2\Phi_2^{\dagger}\Phi_2-
					  \frac{\mu^2}{2}\left(e^{i\phi}\Phi_1^{\dagger}\Phi_2+H.c.\right)+\\
				 &+\frac{\lambda_1}{2}\left(\Phi_1^{\dagger}\Phi_1\right)^2+
				   \frac{\lambda_2}{2}\left(\Phi_2^{\dagger}\Phi_2\right)^2+
				   \lambda_3\left(\Phi_1^{\dagger}\Phi_1\right)\left(\Phi_2^{\dagger}\Phi_2\right)+\\
				 &+\lambda_4\left(\Phi_1^{\dagger}\Phi_2\right)\left(\Phi_2^{\dagger}\Phi_1\right)+
			 	   \frac{\lambda_5}{2}\left[\left(\Phi_1^{\dagger}\Phi_2\right)^2+H.c.\right],
	\end{split}
\end{equation}
where we have used the freedom of field redefinitions to make $\lambda_5$ real. The 2HDM scalar potential, 
Eq.~(\ref{2HDM_potential}), can violate CP either explicitly, via the complex phase $\phi$ in the 
soft $\mathbb{Z}_2$ breaking term $(\Phi_1^{\dagger}\Phi_2+H.c.)$, or spontaneously, due to a relative phase 
between the VEVs of the two doublets. For baryogenesis these extra sources of CP violation are crucial, 
otherwise the predicted BAU will still be far below its measured value. Nevertheless, since we are interested 
in the nature of the electroweak phase transition only, we will restrict ourselves here to a CP conserving scalar 
sector to simplify the analysis, grounding this assumption on previous experience that the phase $\phi$ does 
not influence the phase transition substantially~\cite{Fromme:2006cm}, but keeping in mind that this is just a first 
approach to the problem, and that this limitation should be overcome in future studies.

The doublets and their VEVs at the so-called electroweak minimum can be written as
\begin{equation*}
	\Phi_{i} = \left( \begin{array}{c} \varphi_i^+ \\ h_i+i\eta_i \end{array} \right),
\end{equation*}

\begin{equation}
	 \label{EWmin}
	\langle\Phi_1\rangle = \left( \begin{array}{c} 0 \\ v\cos\beta \end{array} \right),\hspace{0.5cm}
  	\langle\Phi_2\rangle = \left( \begin{array}{c} 0 \\ v\sin\beta \end{array} \right),
\end{equation}
with $v=246/\sqrt{2}$~GeV. The parameter $\beta$, related to the ratio of the VEVs of 
the two doublets, can be interpreted more physically and more conveniently in the following way. From
\[\begin{array}{l} \Phi_1^\prime=\cos\beta\ \Phi_1+\sin\beta\ \Phi_2\\
   \Phi_2^\prime=-\sin\beta\ \Phi_1+\cos\beta\ \Phi_2 \end{array} \implies
   \langle\Phi_1^\prime\rangle= \left( \begin{array}{c} 0\\ v \end{array} \right) \text{ \ and \ } 
\langle\Phi_2^\prime\rangle=0 \]
it becomes clear that $\Phi_1^\prime$ behaves like the SM doublet, so its upper component must be the charged 
Goldstone boson ($G^+$), whereas the lower component contains the neutral Goldstone ($G^0$). Thus, $\beta$ plays 
the role of a \emph{mixing angle} between the charged mass eigenstates ($G^+,H^+$), and also between 
the neutral CP-odd ($G^0,A^0$) ones. We likewise define $\alpha$ to be the 
mixing angle between the lightest and heaviest CP-even fields, denoted $h^0$ and $H^0$. The physical states are then
\begin{alignat*}{2}
	&G^+=\cos\beta\ \varphi_1^+ + \sin\beta\ \varphi_2^+ && \hspace{1cm} \text{(charged Goldstone),}\\
	&H^+=-\sin\beta\ \varphi_1^+ + \cos\beta\ \varphi_2^+ && \hspace{1cm} \text{(charged Higgs),}\\
	&G^0=\cos\beta\ \eta_1 + \sin\beta\ \eta_2 && \hspace{1cm} \text{(neutral Goldstone),}\\
	&A^0=-\sin\beta\ \eta_1 + \cos\beta\ \eta_2 && \hspace{1cm} \text{(CP-odd Higgs),}\\
	&h^0=\cos\alpha\ h_1 + \sin\alpha\ h_2 && \hspace{1cm} \text{(lightest CP-even Higgs)},\\
	&H^0=-\sin\alpha\ h_1 + \cos\alpha\ h_2 && \hspace{1cm} \text{(heaviest CP-even Higgs)}.
\end{alignat*}
Note that our definition of $\alpha$ differs from the general 2HDM literature by an additive factor of $\pi/2$. We 
find our choice more convenient, since the case when $h^0=h_{SM}$ corresponds here to $\alpha=\beta$. In general, 
however, the SM-like Higgs is an admixture of $h^0$ and $H^0$. Note, moreover, that the separation between 
CP-even and CP-odd fields only makes sense because there is no CP violation in the Higgs sector --- otherwise these 
fields would all mix among themselves, and the corresponding mixing angles would have to enter these expressions. 
From this alone one can already appreciate how the assumption of a CP conserving scalar sector simplifies the problem.

The condition that Eq.~(\ref{EWmin}) be indeed a minimum can be used to trade $\mu_1$ and $\mu_2$ for $v$ 
and $\tan\beta$ via
\begin{equation}
	\label{As}\begin{split}
	&\mu_1^2= v^2\left(\lambda_1\cos^2\beta+\lambda_{345}\sin^2\beta\right)-
			    M^2\sin^2\beta,\\
	&\mu_2^2=v^2\left(\lambda_2\sin^2\beta+\lambda_{345}\cos^2\beta\right)-
			   M^2\cos^2\beta,
\end{split}\end{equation}
where $M^2\equiv \mu^2/\sin(2\beta)$ and $\lambda_{345}\equiv\lambda_3+\lambda_4+\lambda_5$. The parameter $M$ plays the role of a natural scale for the masses of the additional scalars, while $h_{SM}$ scales with $v$ as usual. From the diagonalization of the mass matrix we also see that the quartic couplings can be written in terms of the physical parameters as
\begin{equation}\begin{split}
	\label{couplings}
	&\lambda_1=\frac{1}{2v^2\cos^2\beta} 		
		  \left(m_{h^0}^2\cos^2\alpha+m_{H^0}^2\sin^2\alpha-M^2\sin^2\beta\right),\\
	&\lambda_2=\frac{1}{2v^2\sin^2\beta}
		   \left(m_{h^0}^2\sin^2\alpha+m_{H^0}^2\cos^2\alpha-M^2\cos^2\beta\right),\\
	&\lambda_3=\frac{1}{2v^2\sin(2\beta)}
		   \Big[\left(2m_{H^\pm}^2-M^2\right)\sin(2\beta)-\left(m_{H^0}^2-m_{h^0}^2\right)\sin(2\alpha)\Big],\\
	&\lambda_4+\lambda_5=\frac{1}{v^2}\left(M^2-m_{H^\pm}^2\right),\\
	&\lambda_4-\lambda_5=\frac{1}{v^2}\left(m_{A^0}^2-m_{H^\pm}^2\right).
\end{split}\end{equation}
These relations allow us to use as input parameters of the model the physical quantities of the theory, namely 
the masses ($m_{h^0}$, $m_{H^0}$, $m_{A^0}$, $m_{H^\pm}$) and the mixing angles ($\beta$, $\alpha$), as well as 
the only remaining free dimensionful parameter of the potential, the $\mu$ parameter.

\subsection{Electroweak Precision Constraints}

The existence of additional scalar particles running in the loops causes the gauge bosons' two-point functions 
to receive corrections relative to their SM values, so-called ``oblique'' corrections. As a consequence, some 
combinations of gauge boson masses and their couplings, whose experimental values are known to agree with the SM 
prediction to great accuracy, get extra contributions from the new physics introduced. It then becomes a challenge 
for the model to predict a deviation that remains within the precision of the experimental measurement. 
The best example is provided by the $\rho$ parameter, \[ \rho = \frac{m_W^2}{m_Z^2\cos^2\theta_W},\]
which is intimately related to the electroweak symmetry breaking sector of the theory, and whose value is known 
to agree with the SM prediction to better than $0.4$\% at $2\sigma$~\cite{Beringer:1900zz}. Because they contain 
only scalar \emph{doublets}, 2HDMs predict $\rho=1$ at tree-level (as in the SM). 
At loop level, however, there are extra contributions with respect to the SM ones\footnote{In the SM, custodial 
symmetry is violated by $U(1)$ hypercharge gauge interactions and Yukawa couplings, but is preserved by the scalar 
potential.}~\cite{Grimus:2007if}, and one has
\begin{equation}\begin{split}
\Delta\rho^{\mathrm{2HDM}}=\frac{1}{32\pi^2 v^2}&\left[F_{H^{\pm}, A^0}+
\sin^2(\beta-\alpha)(F_{H^{\pm}, h^0} - F_{A^0, h^0})\right.\\
&\left.+\cos^2(\beta-\alpha)\left(F_{H^{\pm}, H^0} - F_{A^0, H^0}\right)\right.\\
&\left.+3\sin^2(\beta-\alpha)\left(F_{Z, H^0} - F_{Z, h^0} - F_{W, H^0} + F_{W, h^0}\right)\right],
\end{split}\end{equation}
with 
\begin{equation}
F_{x,y}= \frac{m_{x}^2+m_{y}^2}{2}-\frac{m_{x}^2\,m_{y}^2}{m_{x}^2-m_{y}^2}\ln\left(\frac{m_{x}^2}{m_{y}^2}\right).
\end{equation}
The condition that $\rho \sim \rho_{SM} \approx 1$ is satisfied only if there is an approximate mass degeneracy 
between the charged and one of the neutral scalars, which is related to the limit in which custodial symmetry is 
recovered~\cite{Haber:2010bw}.

The $\rho$ parameter is only an instance of observables that receive oblique corrections in 2HDMs. For a 
general extension of the SM preserving the $SU(2)_L\times U(1)_{Y}$ gauge structure, these corrections can be 
parametrized by the Peskin-Takeuchi parameters $S$, $T$ and $U$~\cite{Peskin:1991sw} and some higher-order 
extensions of them~\cite{Maksymyk:1993zm}. Nevertheless it turns out that for 2HDMs 
only $\Delta\rho\equiv\rho-1\equiv\alpha_{EM}T$ is relevant, since the experimental bounds on the 
remaining parameters are hardly violated~\cite{Haber:2010bw,Funk:2011ad}.

Another important electroweak precision constraint, unrelated to the oblique parameters above,
comes from $Z\to b\bar{b}$ decays \cite{Haber:1999zh, Jung:2010ik}. We checked explicitly that 
this constraint is milder than the one coming from $B^0-\bar{B}^0$ mixing, 
which we will take into account later (cf. section \ref{section:pheno_constraints}).

\subsection{Finite Temperature Effective Potential}

To study the phase transition we consider the scalar potential of the model at finite temperature, 
which we approximate at 1-loop order, including daisy resummations~\cite{Carrington:1991hz}.
The zero-temperature 1-loop corrections to the potential have the form
\begin{equation}
	V_{\rm loop}=\sum_i \frac{n_i}{64\pi^2}m_i^4\left(\ln \frac{m_i^2}{v^2}-\frac{1}{2}\right),
\end{equation}
with $i$ indexing the particles summed over and $n_i$ their numbers of degrees of freedom. Among the 
fermions we take only the top-quark into account ($n_t=-12$). The other 
fermions can be neglected due to their small masses. Among the bosons we sum over $W^\pm$ ($n_W=6$), 
$Z^0$ ($n_Z=3$), and the scalars. The renormalization scale is taken to be $v=246/\sqrt{2}$ GeV.

Counter-terms ($V_{CT}$) are added so that the zero-temperature 1-loop potential preserves 
the position of the minimum and the masses of the scalar particles. 
Care must be taken at this step, since in Landau gauge the contribution to the scalar masses due to intermediate Goldstone bosons 
running in the loop must be taken into account, and these will be infrared divergent if the total momentum in the loop vanishes. 
This means that renormalizing the scalar masses at $p^2=0$ external momentum is not a well-defined procedure. 
An alternative for renormalizing the Higgs mass on-shell has been developed in Ref. \cite{Cline:1996mga}. Here
we choose to adopt the more straightforward approach of imposing an IR cutoff at $m_{IR}^2=m_{h^0}^2$, which gives a good
approximation to the exact procedure of on-shell renormalization, as argued in \cite{Cline:2011mm}. 

The 1-loop thermal corrections to the effective potential are given by~\cite{Dolan:1973qd}
\begin{equation}
	\label{Vthermal}
	V_{\rm thermal}=\frac{T^4}{2\pi^2}\sum_i n_i\int_0^{\infty} x^2\text{ln}\left(1\mp 
e^{-\sqrt{x^2+m_i^2/T^2}}\right)dx,
\end{equation}
where the sign inside the logarithm is $-$ for bosons and $+$ for fermions. However, evaluating this integral 
numerically is computationally expensive, and it is therefore convenient to introduce an approximate function for it. 
At high temperatures, Eq.~(\ref{Vthermal}) can be approximated by
\begin{equation}\begin{split}
	V^{HT}_{\rm thermal}&\approx T^4\sum_B n_B\left[-\frac{\pi^2}{90}+
						\frac{1}{24}\left(\frac{m_B}{T}\right)^2
						-\frac{1}{12\pi}\left(\frac{m_B}{T}\right)^3
						-\frac{1}{64\pi^2}\left(\frac{m_B}{T}\right)^4\text{ln}\frac{m_B^2}{c_BT^2}\right]\\
			   &+T^4\sum_F n_F\left[-\frac{7\pi^2}{720}
					+\frac{1}{48}\left(\frac{m_F}{T}\right)^2
					+\frac{1}{64\pi^2}\left(\frac{m_F}{T}\right)^4\text{ln}\frac{m_F^2}{c_FT^2}\right]\\
\end{split}\end{equation}
with $c_F=\pi^2\text{exp}\left(\frac{3}{2}-2\gamma\right)$ and $c_B=16c_F$, whereas at low temperatures
\begin{equation}
	V_{\rm thermal}^{LT}\approx-T^4\sum_{i=B,F}n_i\left(\frac{m_i}{2\pi T}\right)^{3/2} \exp\left(-\frac{m_i}{T}\right)\left(1+\frac{15}{8}\frac{T}{m_i}\right).
\end{equation} 
We thus define our approximate 1-loop thermal correction to the effective potential $V_T$ to coincide 
with $V^{HT}$ for $m/T < 1.8$ and with $V^{LT}$ for $m/T>4.5$, being 
a smooth interpolation of these in the region in between such that its first derivative is continuous everywhere. 
Our approximation deviates from the original integral (Eq.~(\ref{Vthermal})) by no more than $4\%$, and for 
most field values the deviation is actually much smaller.

Finally, we also consider thermal corrections to the scalar masses coming from the resummation of daisy 
diagrams~\cite{Carrington:1991hz}. Taking these into account, the mass matrix becomes~\cite{Cline:2011mm}
\begin{equation}
	(M_T)_{ij}=\frac{1}{2}\frac{\partial^2 }{\partial\phi_i\partial\phi_j}
		\left(V_{tree}+\frac{T^2}{24}\sum_i n_i m_i^2\right).
\end{equation}
The thermally corrected masses are then the eigenvalues of $M_T$, and these are the masses that need to be plugged 
into $V_{\rm loop}$ and $V_T$. The final potential is then
\begin{equation}
	V=V_{\rm tree}+V_{\rm loop}+V_{CT}+V_T.
\end{equation}

\section{\label{section:scan}Parameter Scan}

As already mentioned, the parameters we scan over are the physical ones, namely $\beta$, $\alpha$ and 
the scalar masses, together with the $\mu$ parameter. Eqs.~(\ref{As})~and~(\ref{couplings}) are then used 
to get the respective parameters of the potential. We take the mass of the lightest CP-even 
Higgs, $h^0$, to be $m_{h^0}=125$ GeV, i.e. we consider this to be the particle recently discovered at the 
LHC. The other parameters are scanned uniformly over the following ranges:
\begin{center}
	$0.4\leq\ \tan \beta\leq 10$,\\
	$-\frac{\pi}{2}<\ \alpha\ \leq\frac{\pi}{2}$,\\
	$0\text{ GeV}\leq\ \mu\leq 1\text{ TeV}$,\\
	$100\text{ GeV}\leq\ \ m_{A^0},\ m_{H^\pm}\leq 1\text{ TeV}$,\\
	$150\text{ GeV}\leq\ m_{H^0}\leq 1 \text{ TeV}$.\\
\end{center}

From these randomly generated points in the parameter space, we exclude those that do not pass the 
EW precision tests at $2\sigma$~\cite{Beringer:1900zz} 
or the condition\footnote{Here, perturbativity is imposed at the 
electroweak scale $v$ (as in~\cite{Drozd:2012vf}). Alternatively, one can impose perturbativity all the way up to 
the cut-off scale of the model~\cite{Grinstein:2013npa}.} $\lambda_{1-5}<4\pi$. Next, we check if the EW 
minimum is at least metastable\footnote{We use the word ``metastable'' meaning metastability with a
long life time.} by randomly searching for other minima in a region of $1$ TeV radius around it. If any 
other minimum is found to be deeper than the EW one, the point is discarded. Note, in particular, 
that we do \emph{not} impose the conditions that the tree-level potential be bounded 
from below --- what matters, after all, is the loop corrected potential. If all these tests are 
passed, the point is said to be \emph{physical}, and we proceed to evaluate the phase transition.

This we do by increasing the temperature by small steps and following the minimum of the potential 
(starting from the EW minimum at zero temperature), until the potential at this minimum overcomes 
its value at the origin, or a certain maximum temperature is reached above which we do not expect to get 
a significant number of points with strong phase transition. Here we take $T_{\rm step}=6$ GeV 
and $T_{\rm max}=300$ GeV, and we find that less than $0.5$\% of the points with a strong 
phase transition have $T_c>200$ GeV. 
For greater precision, if a phase transition is found we repeat the process with $T_{\rm step}=1$ GeV, 
starting from the last minimum found to be below the origin.

The critical point is taken to be the last one for which the minimum lied below the origin, 
and the strength of the phase transition is then calculated as the ratio between the norm of the VEV and the 
temperature at this point\footnote{Note that this ratio, even though being used as a standard measure
of the strength of the electroweak phase transition, suffers from problems related to 
gauge invariance~\cite{Dolan:1973qd,Patel:2011th}. An appropriate, gauge invariant definition of the $\xi$ parameter that measures the strength of the electroweak 
phase transition has been given in~\cite{Patel:2011th,Wainwright:2011qy} (see also~\cite{Garny:2012cg}). }, 
\[ \xi = \frac{v_c}{T_c}\ .\]
The phase transition is considered to be strong if $\xi>1$. This ensures that a baryon number generated 
during the phase transition is not washed out afterwards~\cite{Moore:1998swa}. A point in the parameter 
space for which this condition is satisfied will be called a ``strong PT point''.

Of course, there is the possibility that we are overcounting the number of strong PT points by choosing
$T_{\rm step}=1$~GeV and not smaller. To estimate how often that will be the case, we also calculate the phase transition
strength at the next step, i.e. at $T=T_c+1$~GeV. If $\xi(T_c)>1$ but $\xi(T_c+1~\text{GeV})<1$ then this may actually 
not be a strong PT point, even though we consider it to be. We find that in our case this occurs for no more than $5$\% of the sample.

Out of the approximately $6.3\times 10^6$ points initially scanned, about $81$\% are discarded already 
from EW precision 
tests. This huge waste of points is due to the complete randomness of our scan, in particular due to 
its picking the masses of all scalar particles independently. Indeed, as mentioned above, only points with 
an approximate mass degeneracy are expected to pass the $\rho$ parameter test. A natural criterion for 
this approximation is that $m_{H^\pm}$ lie in an interval of size about $v$ around the mass of some other 
scalar, in which case only a fraction of $v/(900~{\rm GeV})\sim 19$\% of points are expected to 
survive\footnote{The $900$ GeV in the denominator being the range of the scan over $m_{H^\pm}$.}, in good 
agreement with our actual findings. 

Yet another $17$\% of the initial sample is discarded due to the couplings being larger than the bound 
required by perturbativity, and this also stems from the random nature of our scan. Indeed, since $v$ 
and $\mu$ are the only free dimensionful parameters of the model, they act as natural scales for the 
scalar masses, and the role of the quartic couplings is, roughly speaking, to regulate the deviation of 
these masses from these base values as well as their splitting among 
themselves --- cf. Eq.~(\ref{couplings}). These couplings are thus expected to be numerically large in a scan 
where the values of $\mu$, $\tan\beta$ and the masses are all chosen independently. 

Among the surviving points, about $18$\% pass the metastability test. Hence, 
only $0.41$\% of the initial sample are \emph{physical} points which are tested for phase transition. 
Nevetheless, because the total number of points scanned was about $6.3\times 10^6$, the number of physical 
points is still about $2.6\times 10^4$. Finally, about $16.5$\% of these physical points have a strong phase 
transition, so we end up with about $4.3\times 10^3$ points, which is large enough to provide significant 
statistics concerning the general behaviour of the electroweak phase transition 
with respect to the input parameters, as will be shown in Section~\ref{section:data}.

Table~\ref{tab:scan} summarizes the discussion of this section.
\begin{table}
   \centering
   \begin{tabular}{|c |c|c| c| c| c|}
      \hline
     \ & Total & EW precision & $\lambda_i<4\pi$ & Metastability & Strong PT \\ \hline
      Absolute & $6.3\times 10^6$ & $1.2 \times 10^6$ & $1.4\times 10^5$ & $2.6\times 10^4$ & $4.3\times 
10^3$ \\ \hline
	Relative & 100\% & 19.1\% & 2.3\% & 0.41\% & 0.069\% \\ \hline
   \end{tabular}
   \caption{\small{Number of points of the initial sample that survive after each step of tests.}}
   \label{tab:scan}
\end{table}

\section{\label{section:data}Analysis and Results}

\subsection{\label{section:pheno_constraints}Constraints from Flavour Physics}

In performing our parameter scan we were able to avoid specifying the 2HDM-Type, since both the physical constraints imposed 
(vacuum metastability, perturbativity and electroweak precision constraints) and the dynamics of the phase transition 
do not depend on the specific type of 2HDM under consideration. 

However, a specific choice for the Yukawa couplings' pattern has to be made in order to confront 
the results of our scan with both constraints from flavour physics and LHC data. Focusing at this stage 
on flavour observables, for 2HDMs (with a $\mathbb{Z}_2$-symmetry at most softly broken) with parameters 
in our range of scan, the only relevant constraints come from $\bar{B}\to X_s\gamma$ decays and 
$B^0-\bar{B}^0$ mixing~\cite{Mahmoudi:2009zx}. Because leptons play 
no role in these effects, these constraints are the same for Types~I and X and for Types~II and Y (cf. 
Table~\ref{tab:Types}), which means we can consider these pairs as indistinguishable.

Constraints from $B^0-\bar{B}^0$ mixing are type-independent and exclude very low values of $\tan\beta$. As 
for $\bar{B}\to X_s\gamma$ decay constraints, they affect Type~I/X models by restricting lower values 
of $\tan\beta$ even more severely, whereas in Type~II/Y they imply a lower bound for the mass of the charged 
scalar~\cite{Hermann:2012fc}, 

\[ m_{H^\pm} \ge 360~\text{GeV at } 95\% \text{ CL}. \]

\begin{figure}[h]
\centering
   \includegraphics{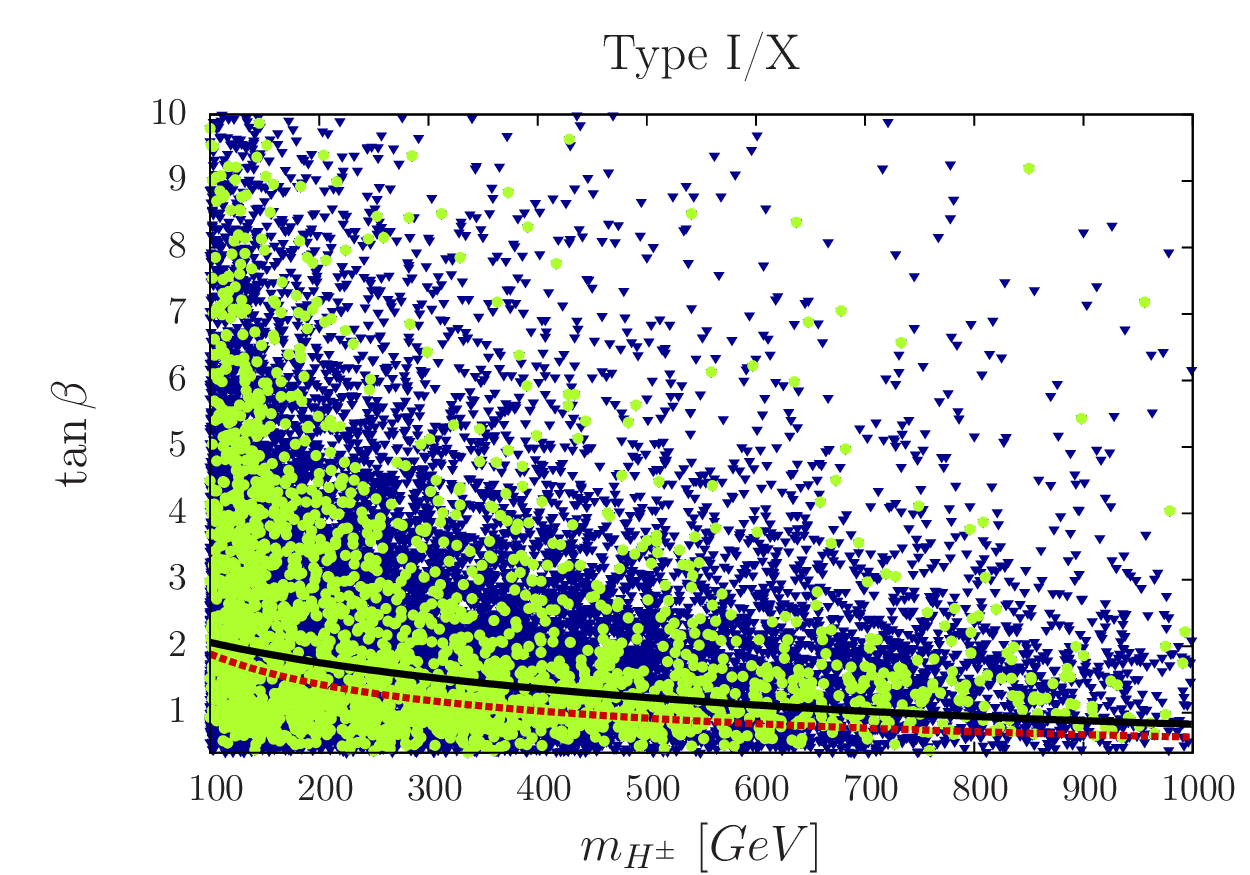} 
   \caption{\small{Scatter plot in $\tan\beta\times m_{H^\pm}$ showing the exclusion regions 
from $\bar{B}\to X_s\gamma$ (red/dashed) and $B^0-\bar{B}^0$ mixing (black/full) for Types I/X. Blue/dark-grey 
points are physical, while green/light-grey ones have a strong phase transition.}}
   \label{fig:scatter_plot1}
\end{figure}

\begin{figure}[h]
\centering
   \includegraphics{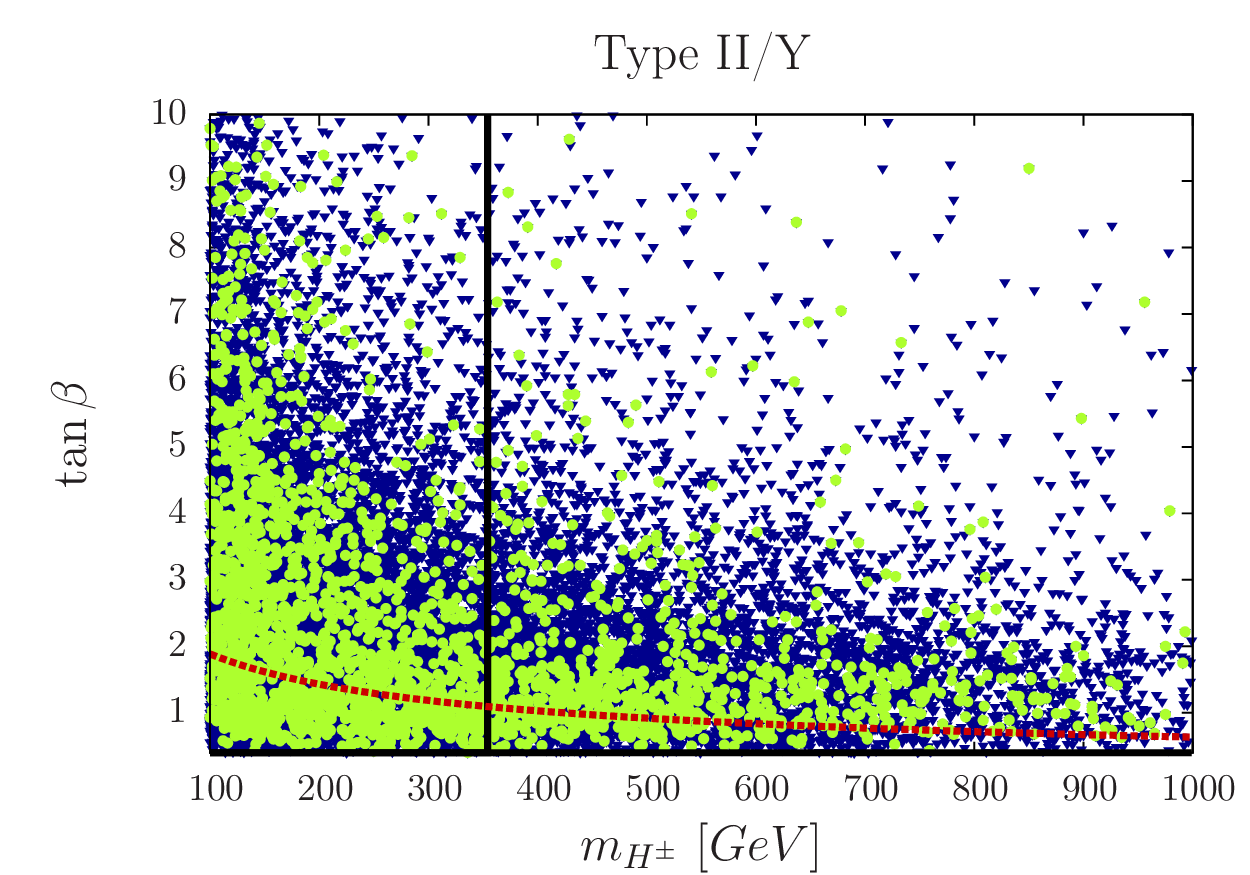}
   \caption{\small{Scatter plot in $\tan\beta\times m_{H^\pm}$ showing 
the exclusion regions from $\bar{B}\to X_s\gamma$ (red/dashed) and $B^0-\bar{B}^0$ mixing (black/full) for 
Types II/Y. 
Blue/dark-grey points are physical, while green/light-grey ones have a strong phase transition.}}
   \label{fig:scatter_plot2}
\end{figure}

Figures~\ref{fig:scatter_plot1}~and~\ref{fig:scatter_plot2} show how the data of our scan are distributed 
in the $(m_{H^\pm},\tan\beta)$ plane, 
as well as the exclusion curves from $\bar{B}\to X_s\gamma$ and $B^0-\bar{B}^0$ mixing for Types~I/X and II/Y. 
The most severe cut occurs for Types~II/Y, where only $34\%$ of our original data sample survives, 
while $65\%$ of our data sample passes the flavour constraints for Types~I/X. 

\subsection{\label{section:data2}Confronting the EW Phase Transition with LHC Data}

We now analyze our data sample and show the regions of the parameter space in 2HDMs that favour 
a strongly first-order EW phase transition. At the same time, we will confront those regions with 
the recent data from the $7$ and $8$ TeV run of LHC 
(see~\cite{Grinstein:2013npa,Celis:2013rcs,Krawczyk:2013gia,Barroso:2013zxa,Shu:2013uua, Chen:2013rba,Craig:2013hca,Eberhardt:2013uba} 
for recent analyses of 2HDMs in the 
context of LHC results). The results will be presented as histograms 
with the number of physical (blue/dark-grey) and strong PT (green/light-grey) points as a function of 
a given parameter (the green/light-grey bars will always be rescaled by a factor of 2, for convenience). More important
than the actual distribution shown in these histograms, which will depend heavily on the particular scanning method we choose,
is the ratio 
\[ \mathcal{P}_{\xi>1} \equiv \frac{\text{\# strong PT points}}{\text{\# physical points}}~, \]
indicating the probability of having a strong phase transition (PT) as a function of the parameter under consideration. This quantity
will be plotted in solid lines. Still, the actual distribution of the counting rates is important, especially because $\mathcal{P}_{\xi>1}$ becomes a less
precise indicative of that probability, the smaller the number of physical points in a given range.

Before jumping into the data analysis itself, a brief discussion on the scale structure of the model shall 
prove very enlightening. Note that the full Lagrangian of the electroweak theory with two scalar doublets 
contains three dimensionful parameters, namely $\mu_1$, $\mu_2$ and $\mu$. The values of these parameters, 
therefore, determine the energy scale of the resulting theory. In particular, the position of the minimum of 
the potential scales as 
\[ v^2 \sim \mu_1^2 + \mu_2^2 + \frac{\mu^2}{\sin(2\beta)}, \]
as can be seen from Eq.~(\ref{As}), so we expect $\mu_1,\mu_2,\mu \sim 10^2$ GeV. Of course, in practice one 
can fix $v$ and take $\mu$ arbitrarily, tuning the remaining parameters so that Eq.~(\ref{As}) is satisfied, 
thus guaranteeing the existence of a minimum at the desired scale. But it turns out that, if $\mu\gg v$, 
this minimum will be so unnatural that the potential will generally end up with a second, \emph{deeper} one 
at $O(\mu)$, and the artificially created minimum will not be metastable. 

The same reasoning also leads to a preferred $\sin(2\beta)\sim 1$ (and the larger 
$\mu$ is, the more so), implying $\tan\beta\sim 1$. This can also be seen from the fact that $\mu_1$ 
increases with $\tan\beta$, whereas $\mu_2$ goes roughly as $(\tan\beta)^{-1}$. Since we expect that 
$\mu_1\sim\mu_2\sim v$, then $\tan\beta$ cannot be neither too large nor too small, and in turn $\tan\beta\sim 1$ 
is favoured.

\subsubsection{Dependence on $\mu$, $\tan\beta$ and $\alpha$}

\begin{figure}[h!]
	\centering
	   \includegraphics[scale=0.6]{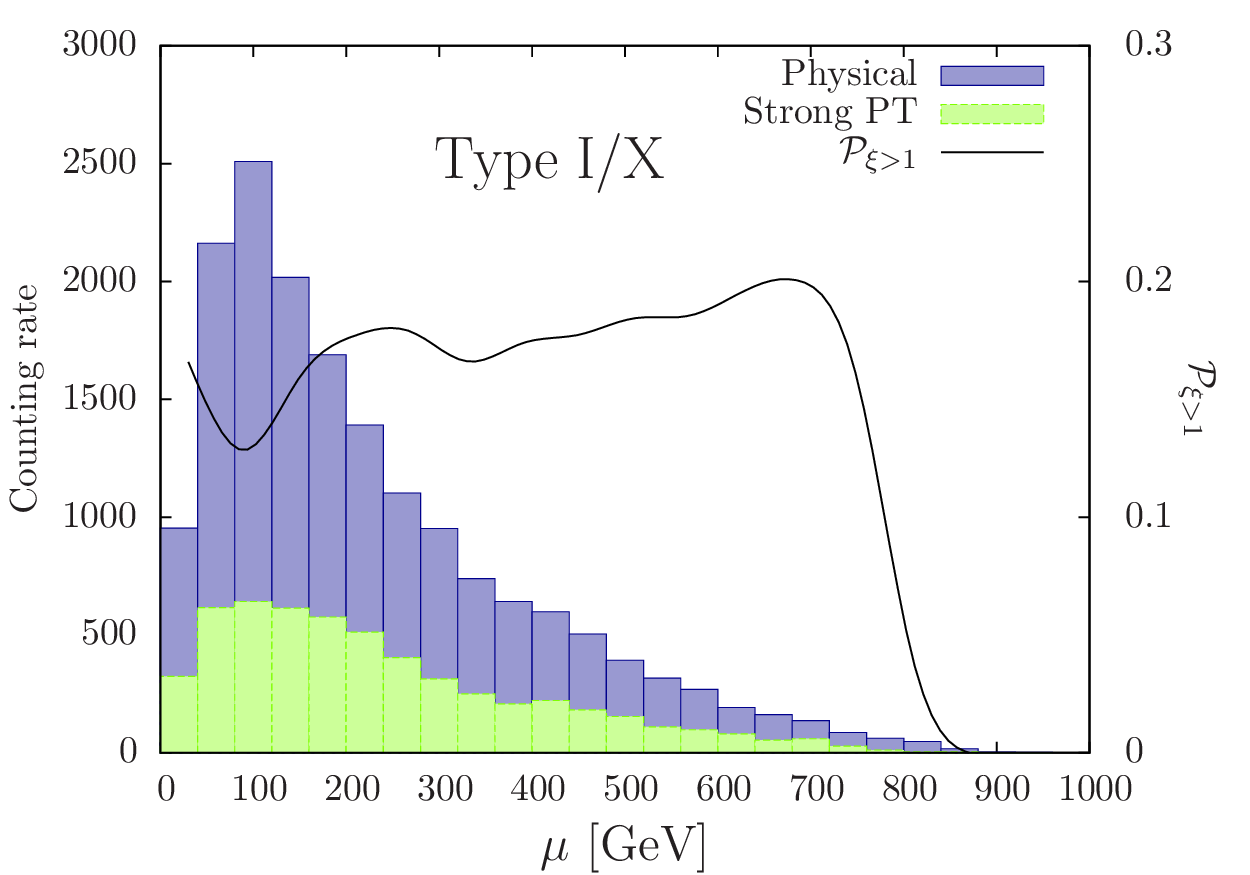}
	   \includegraphics[scale=0.6]{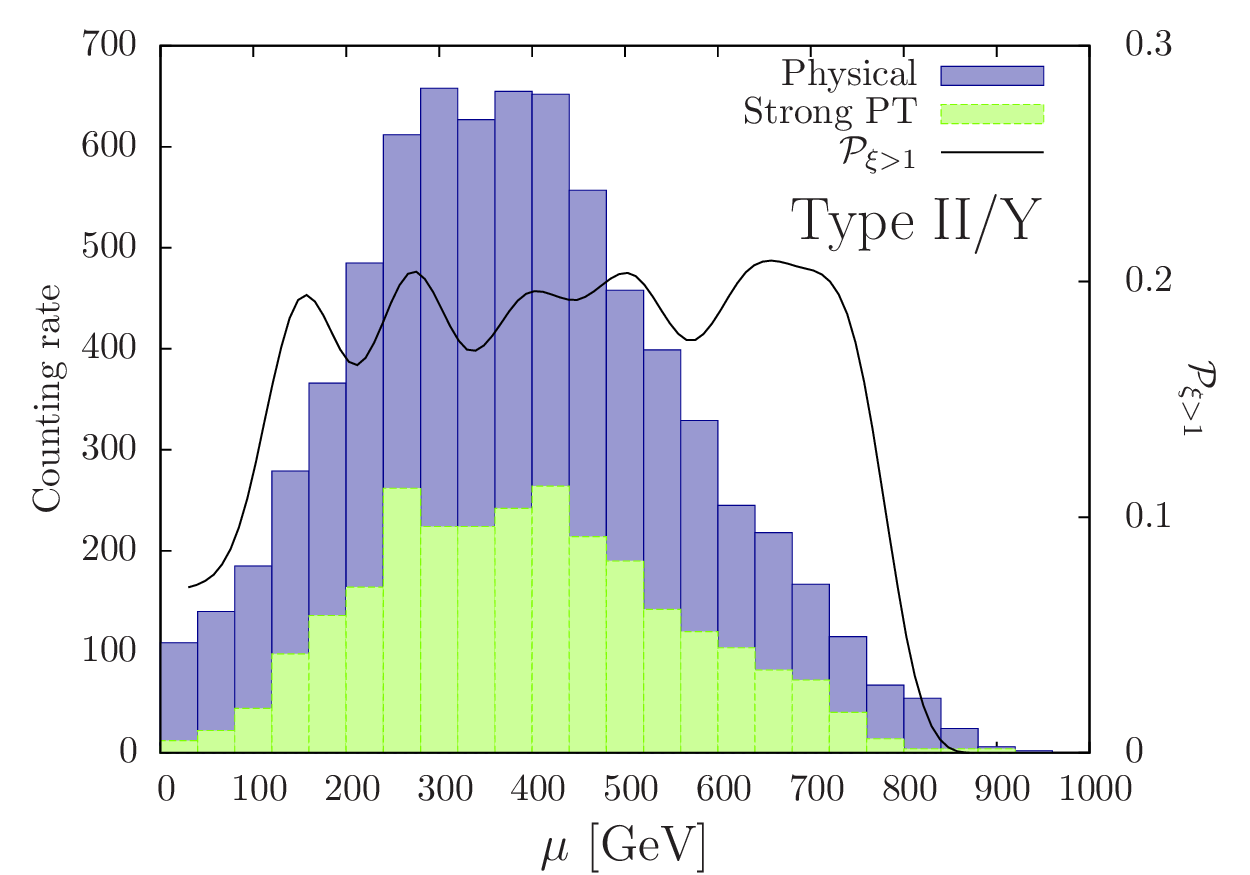}\\[5mm]

	   \includegraphics[scale=0.6]{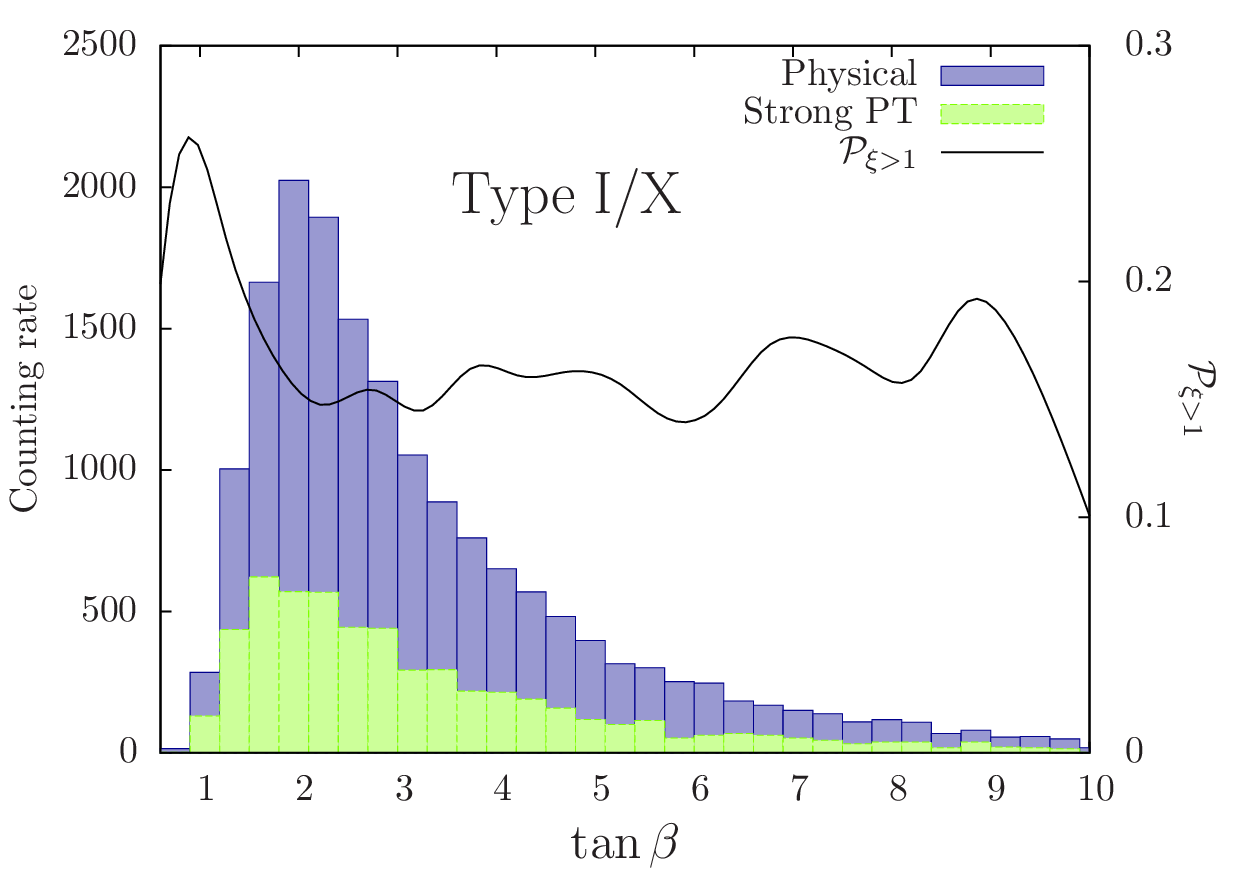}
	   \includegraphics[scale=0.6]{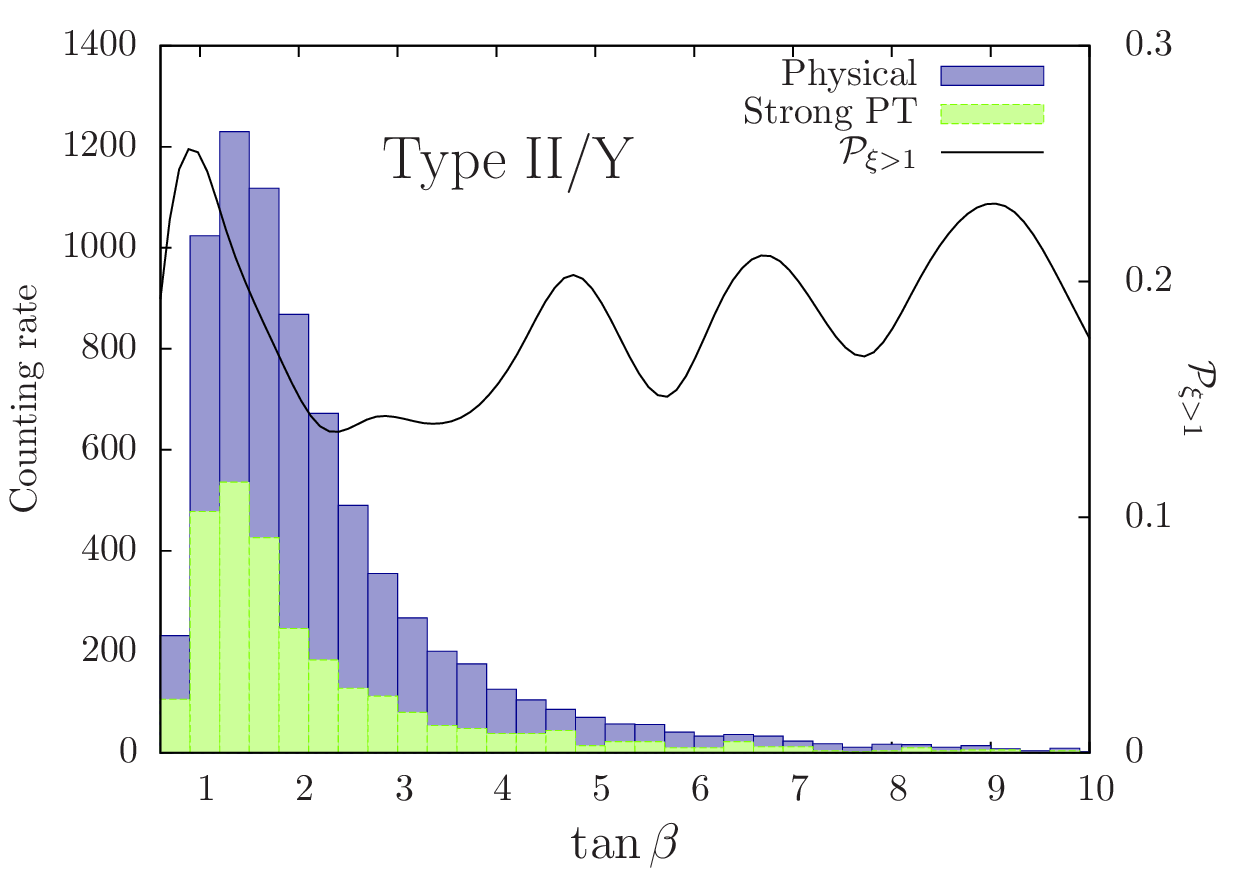}

	\caption{\small{Counting rates for physical (blue/dark) and strong phase transition (green/light) points, and their ratio, as a function 
of $\mu$ (top) and $\tan\beta$ (bottom) for Types I/X (left) and II/Y (right).}}
	\label{fig:mu_tanb}
\end{figure}

The expectations outlined above are confirmed by the plots in Fig.~\ref{fig:mu_tanb}. The lower bound 
of $m_{H^\pm} \ge 360$~GeV in Types~II/Y is reflected in a preference for larger values of $\mu$, which in turn 
results in a sharpening of the peak around $\tan\beta\approx 1$. On the other hand, the constraints on 
Types~I/X disfavour low values of $\tan\beta$, so the peak is slightly displaced towards $\tan\beta\approx 2$. 
Large values of $\tan\beta$ are also disfavoured by our choice of scanning over the physical parameters, 
rather than over the couplings. This is because $\lambda_1$ and $\lambda_3$ grow with $\tan\beta$ (cf. Eq. (\ref{couplings})),
so for large $\tan\beta$ the perturbativity requirement will only be satisfied if the other parameters are tuned to counter-balance this growth, 
which will rarely be the case when they are scanned randomly.

From Fig.~\ref{fig:mu_tanb} we note, moreover, that $\tan\beta\approx 1$ is even more favoured by the 
requirement of a strong phase transition\footnote{The highly oscillatory behaviour of $\mathcal{P}_{\xi>1}$ for $\tan\beta\gtrapprox 5$, 
particularly for Types II/X, is a consequence of the low luminosity in this region, mostly due to the tuning refered to in the previous
paragraph.}. This is an excellent result from the baryogenesis perspective, 
since the effective CP violation coming from 2HDMs decreases with $\tan\beta$, so that the net baryon 
number generated is expected to be suppressed as $n_B\sim (\tan\beta)^{-2}$~\cite{Huber:2000mg}. 
If it were the case that a strong phase transition preferred larger values of $\tan\beta$, then 
it would be harder to generate a baryon asymmetry since an increase of the phase transition strength would cause a 
suppression of CP violation and vice-versa. But the very opposite occurs, and 2HDMs prefer the situation 
that is most favourable for baryogenesis.

It is also worth pointing out that, for $\mu \gtrsim 1$ TeV, one can barely find a point that yields a 
physically acceptable theory, i.e. with a metastable EW minimum. This explicitates that fine-tuning is 
needed if a light scalar is embedded in a theory with a large mass scale.

Another interesting result concerns the mixing angle $\alpha$ that regulates the couplings of the CP-even neutral 
scalars to the fermions. In our scan we assume the lightest of these scalars, $h^0$, to be the bosonic resonance 
recently found at LHC. The current experimental data can not confirm yet whether this discovered 
particle behaves as the SM Higgs, but its decay rates seem to roughly agree with the SM predictions. 
Thus, $\alpha\approx\beta$ seems to be experimentally preferred. In fact, the LHC measurement of the 
coupling of $h^0$ to $W$ and $Z$ bosons imposes the constraint (see for example~\cite{Falkowski:2013dza})

\[ \mathrm{cos}(\alpha-\beta) > 0.7 \quad \mathrm{at}\, 95\% \, \mathrm{C.L.}\]

\begin{figure}[h!]
	\centering
	\includegraphics[scale=0.6]{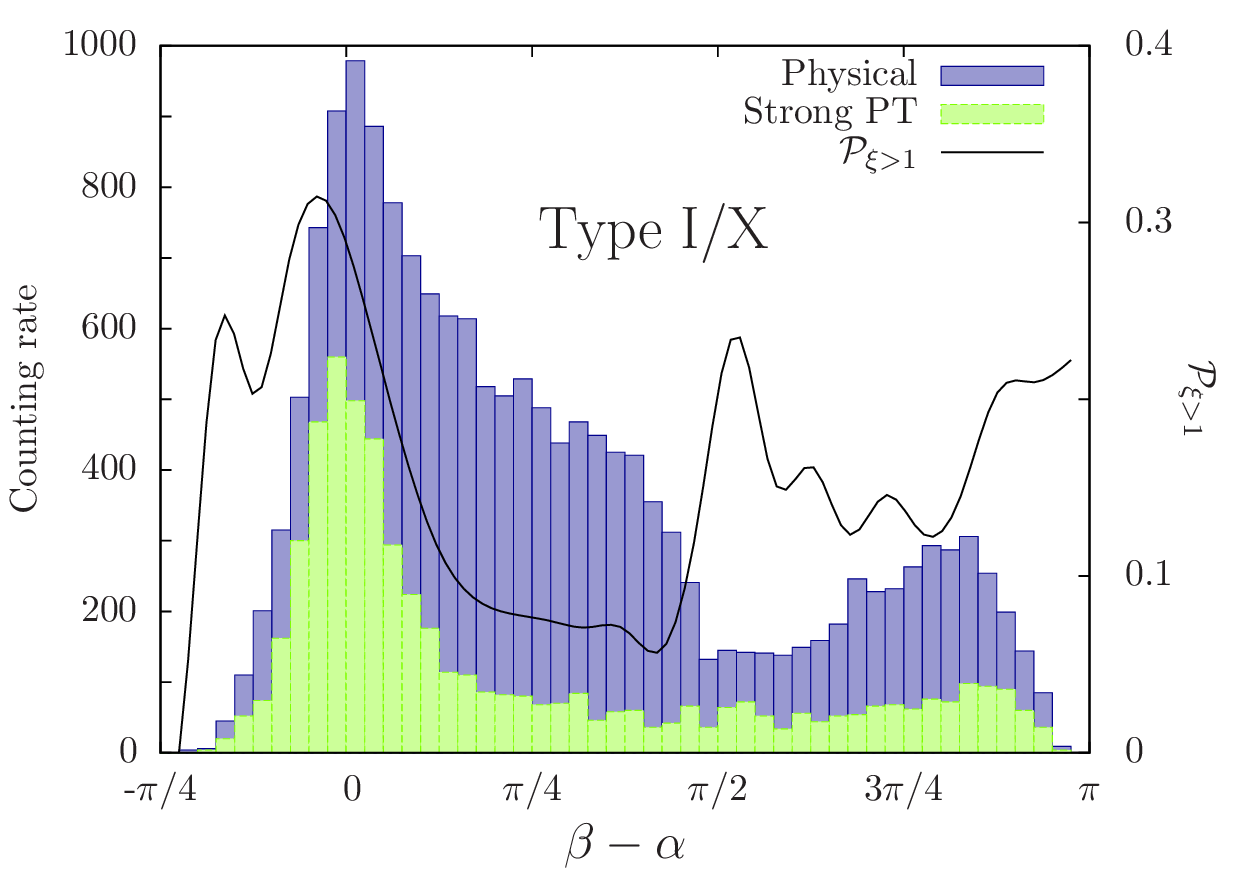}
	\includegraphics[scale=0.6]{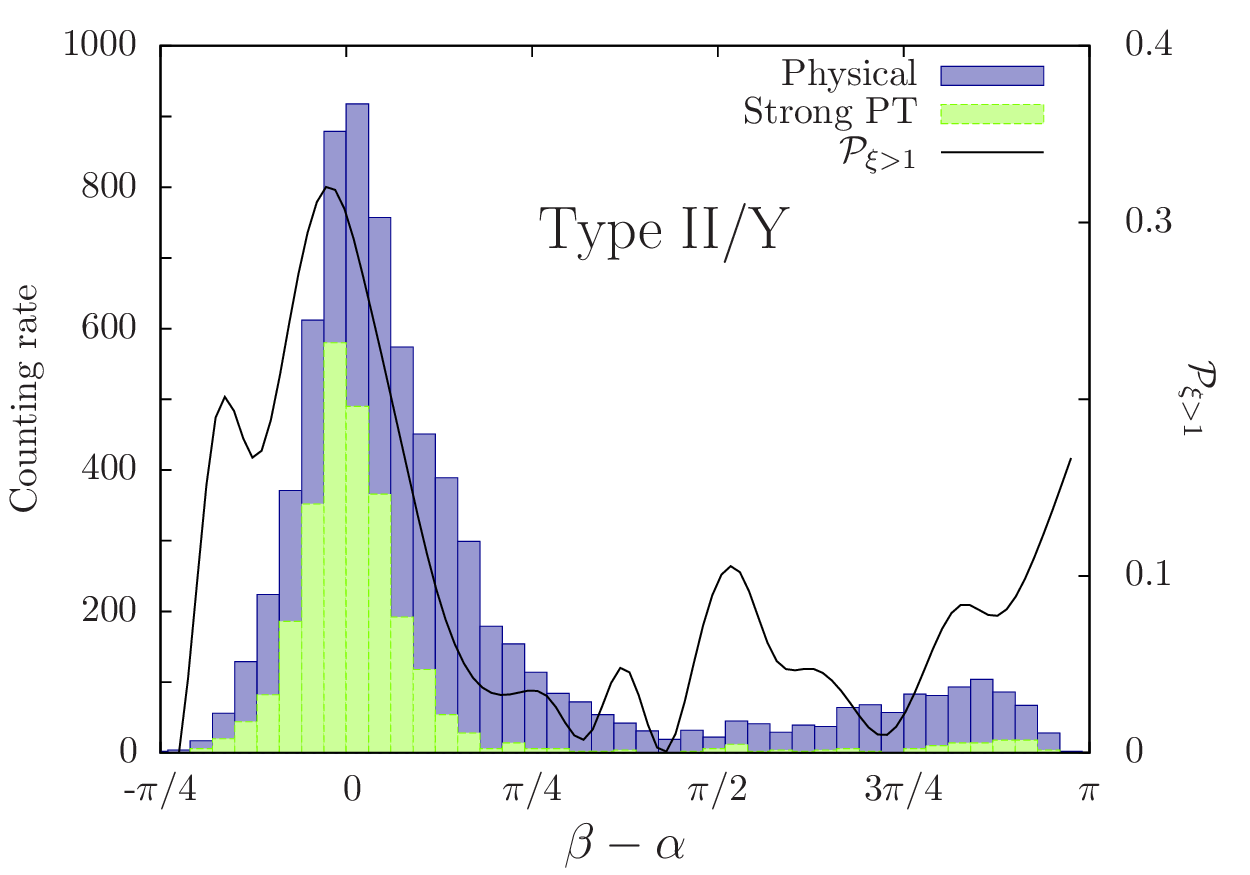}
	\caption{\small{Counting rates for physical (blue/dark) and strong phase transition (green/light) points, and their ratio, as a function 
of $\beta-\alpha$ for Types~I/X (left) and Types~II/Y (right).}}
	\label{fig:b-a}
\end{figure}

Fig.~\ref{fig:b-a} shows that this is also the preferred behaviour in 2HDMs, and even more so when we require a strongly first-order phase transition. 
This is even more pronounced in Type~II/Y models, as a consequence of their favouring larger masses. Finally, 
we can see that for strong phase transitions there is a slight preference for $\alpha \lesssim \beta$, which means 
the couplings of $h^0$ (compared to $h_{SM}$) to up-type quarks are slightly enhanced (and a slightly enhanced 
production cross section at LHC for $h^0$ w.r.t $h_{SM}$ is then preferred). The way this affects down-type 
quarks and leptons depends on the particular 2HDM type one is dealing with, and this will play a role when we look 
at the branching ratios for $h^0\to\gamma\gamma$.

\subsubsection{Masses and couplings}

\begin{figure}[h!]
	\centering
	\includegraphics[scale=0.6]{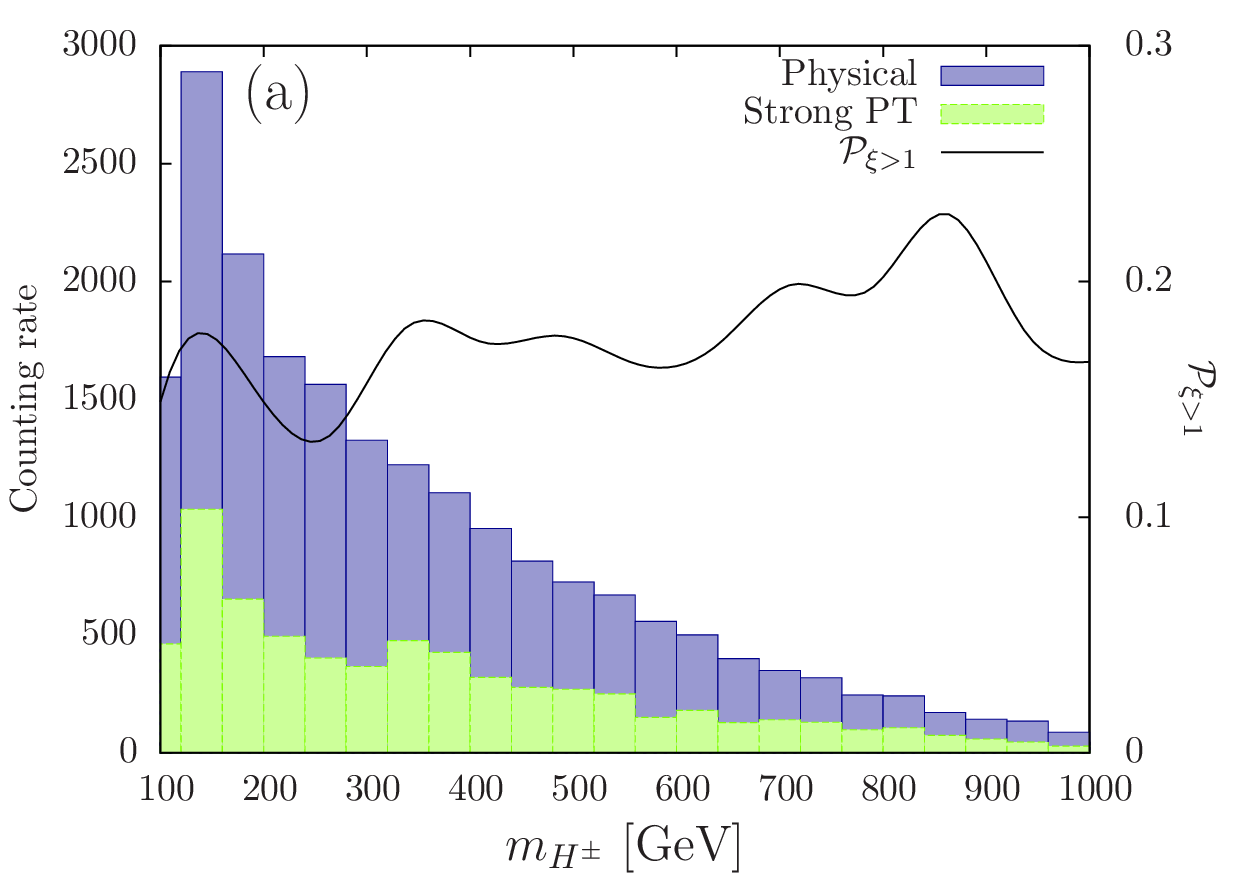}
	\includegraphics[scale=0.6]{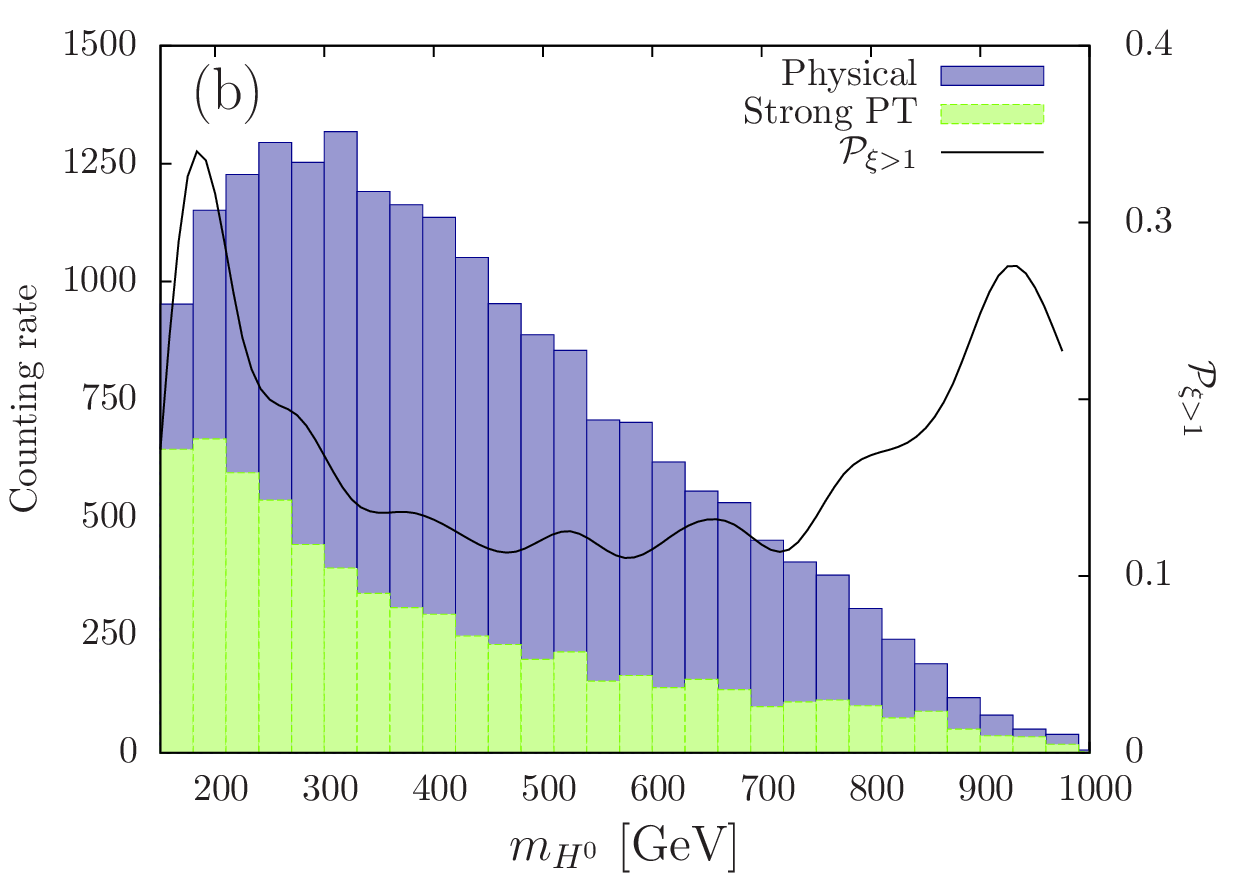}\\[5mm]
	\includegraphics[scale=0.6]{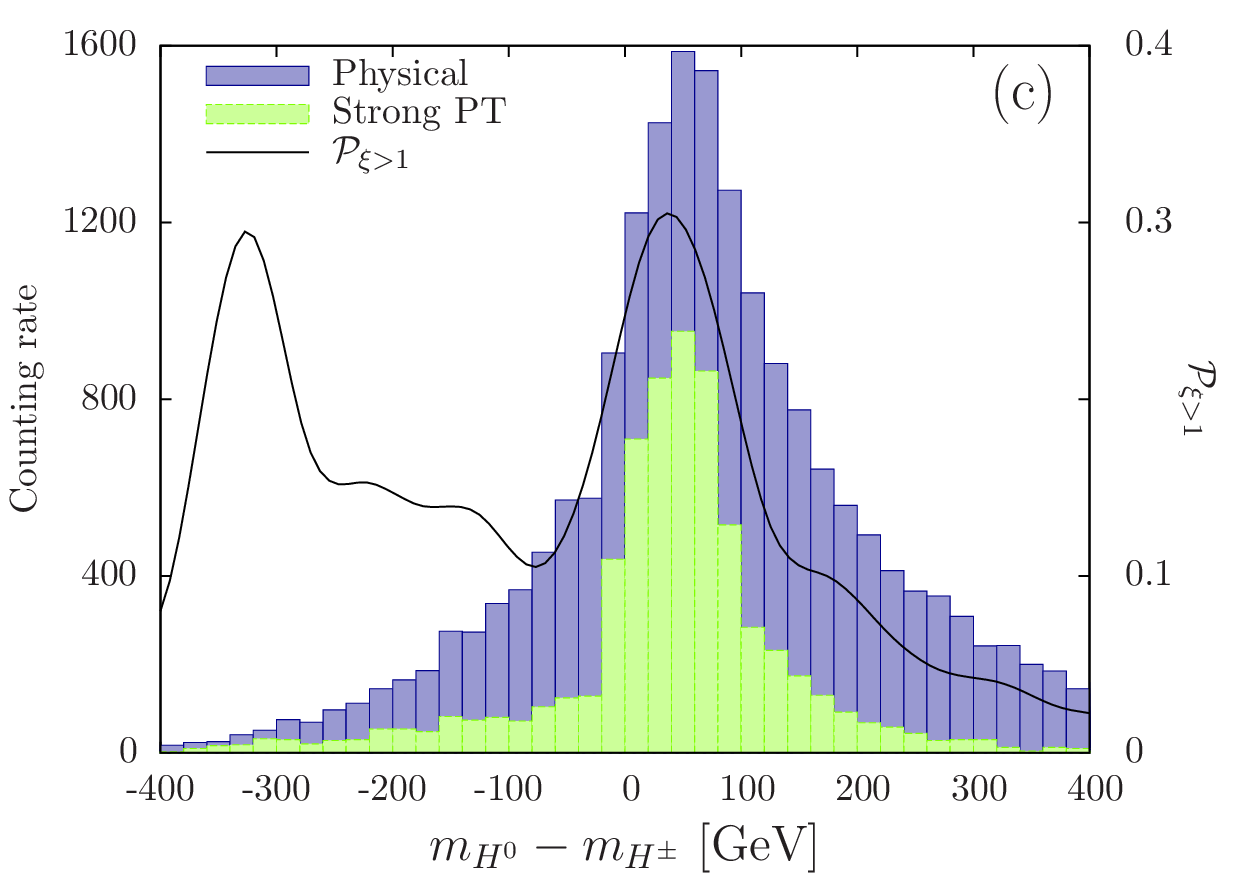}
	\includegraphics[scale=0.6]{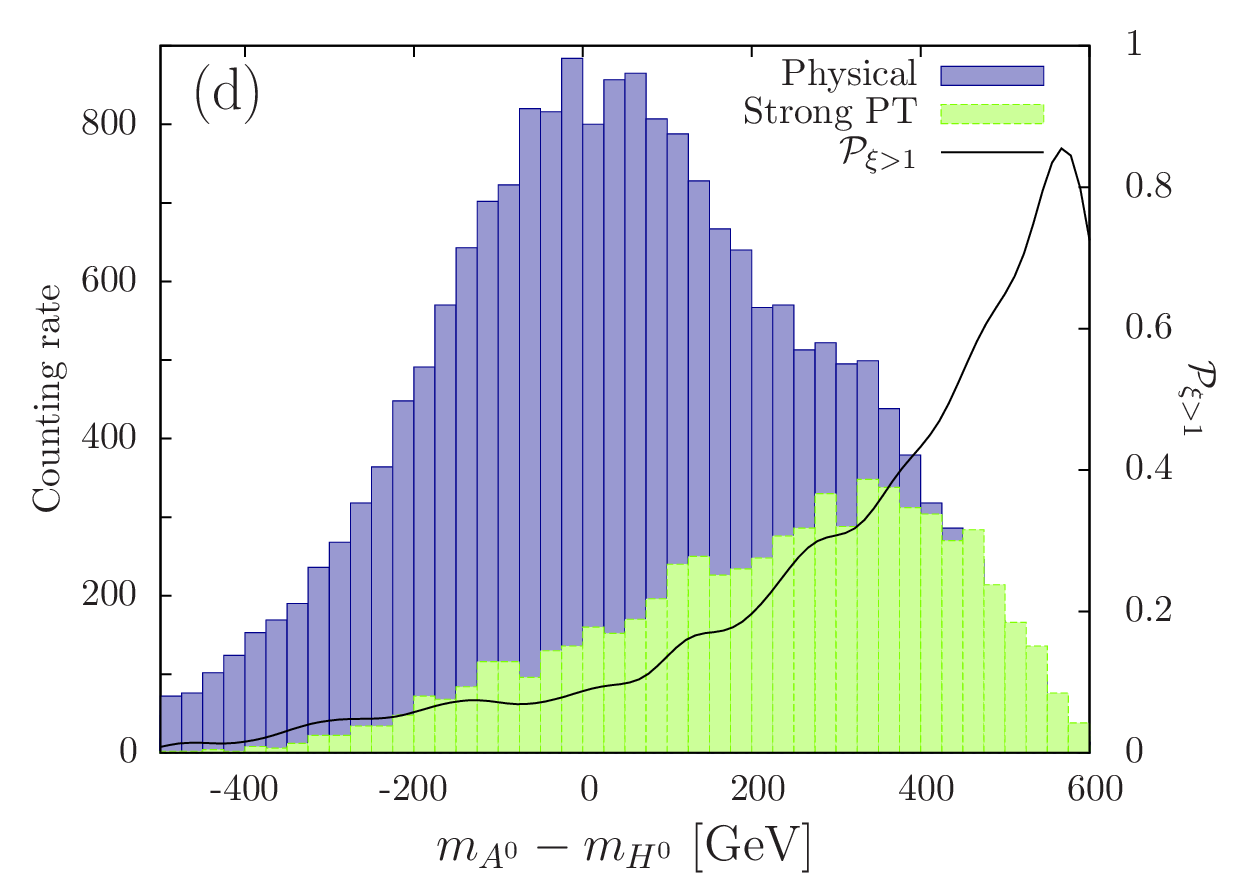}\\[5mm]
	\includegraphics[scale=0.6]{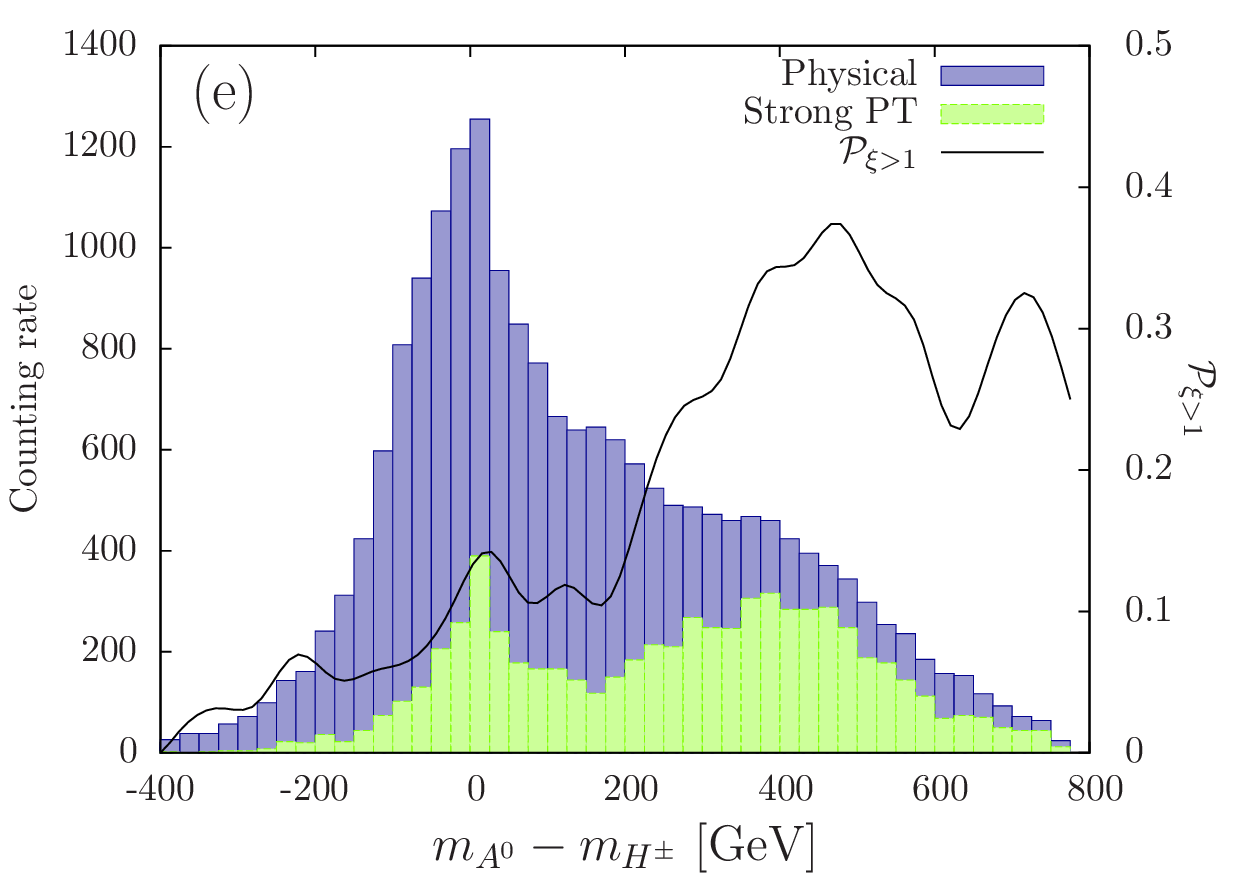}
	\includegraphics[scale=0.6]{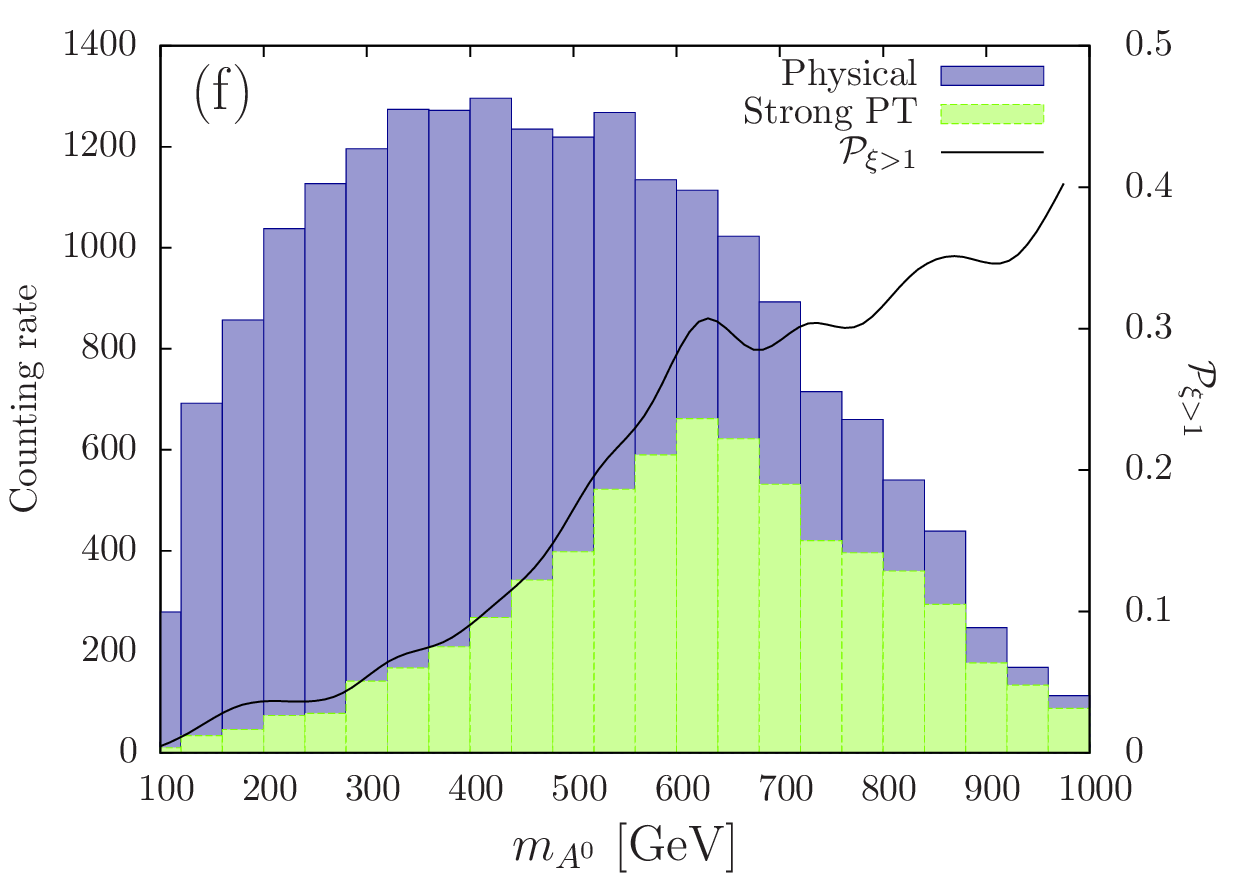}\\[5mm]
	\caption{\small{Counting rates and ratio for points subject only to type-independent constraints 
from $B^0-\bar{B}^0$ mixing. (a) $m_{H^\pm}$ hardly influences the phase transition. (b) Preference for $m_{H^0}\approx 200$ GeV; (c--e) Strong phase transitions 
prefer a scalar mass hierarchy $m_{A^0}>m_{H^0}\gtrsim m_{H^\pm}$. (f) Large pseudo-scalar masses, $m_{A^0}\gtrsim 400$ GeV, are also favoured.}}
	\label{fig:tendencies}
\end{figure}

As for the influence of the masses on the phase transition, the results show four tendencies that are 
independent of the model type considered: (i) $m_{H^\pm}$ is hardly 
influential; (ii) $m_{H^0}\approx 200$ GeV; (iii) $m_{A^0}> m_{H^0} \gtrsim m_{H^\pm}$ is a preferred hierarchy; (iv) $m_{A^0}\gtrsim 400$ GeV. 
These assertions are supported by Fig.~\ref{fig:tendencies}, where only the model-independent constraint 
from $B^0-\bar{B}^0$ mixing is taken into account, so that the plots reflect a general behaviour of phase 
transitions in 2HDMs. 

\begin{figure}[h!]
	\centering
	\includegraphics[scale=0.6]{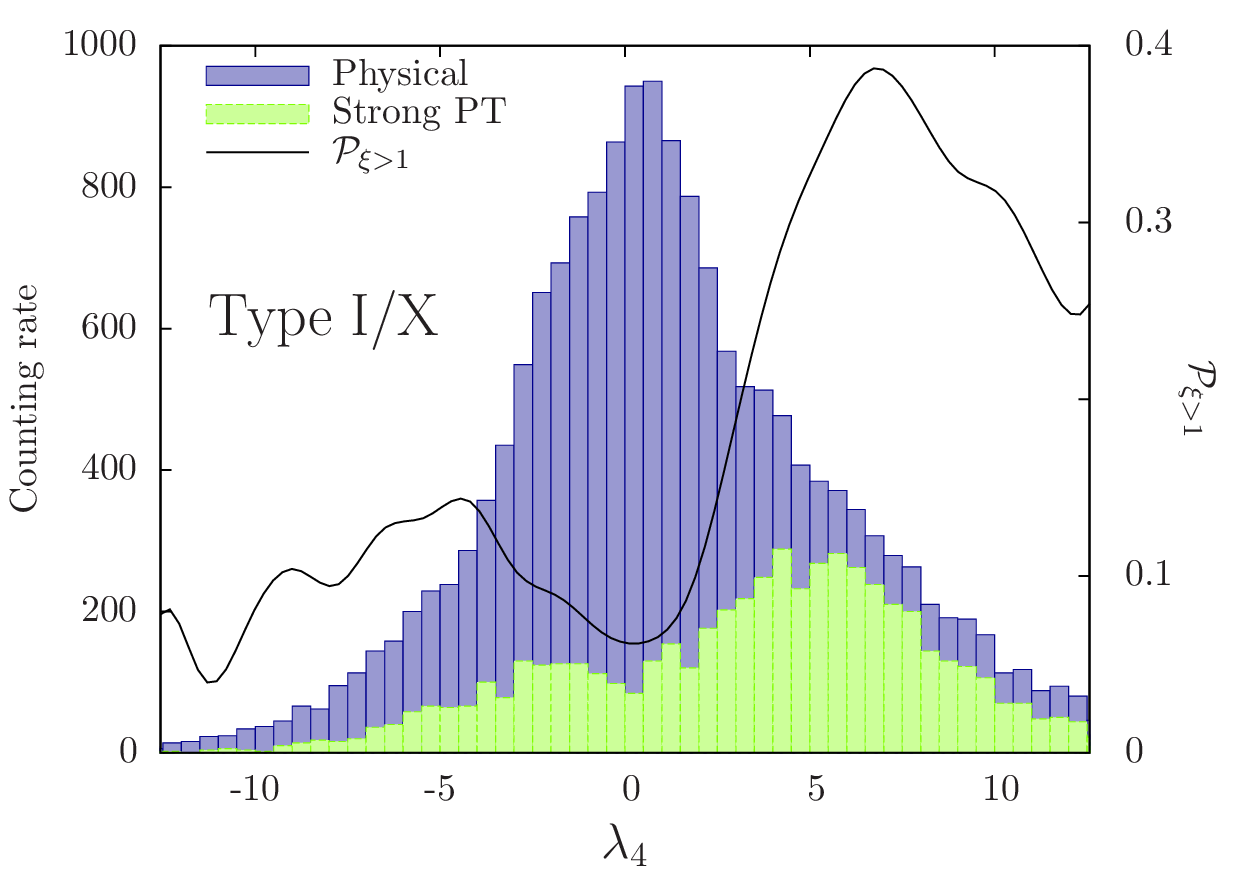}
	\includegraphics[scale=0.6]{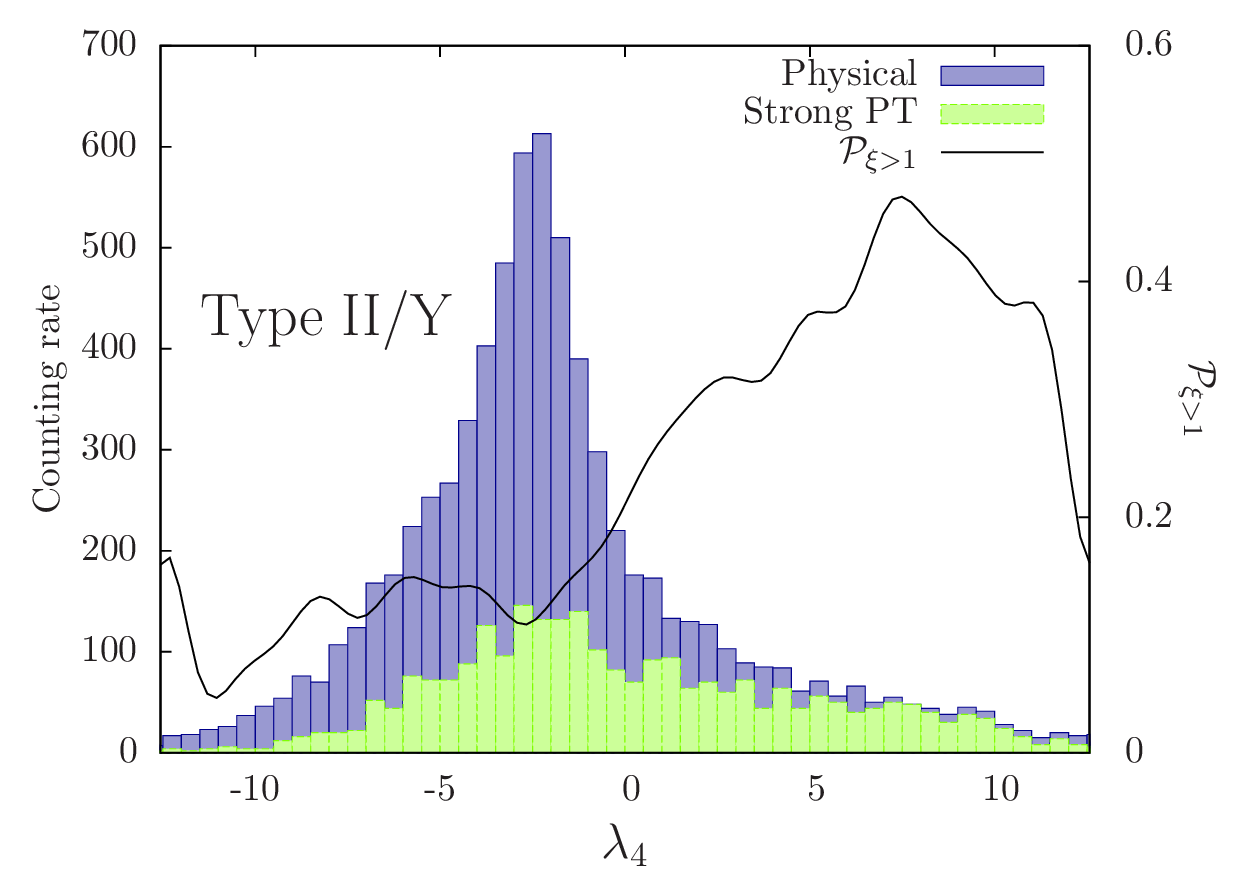}\\[5mm]
	\includegraphics[scale=0.6]{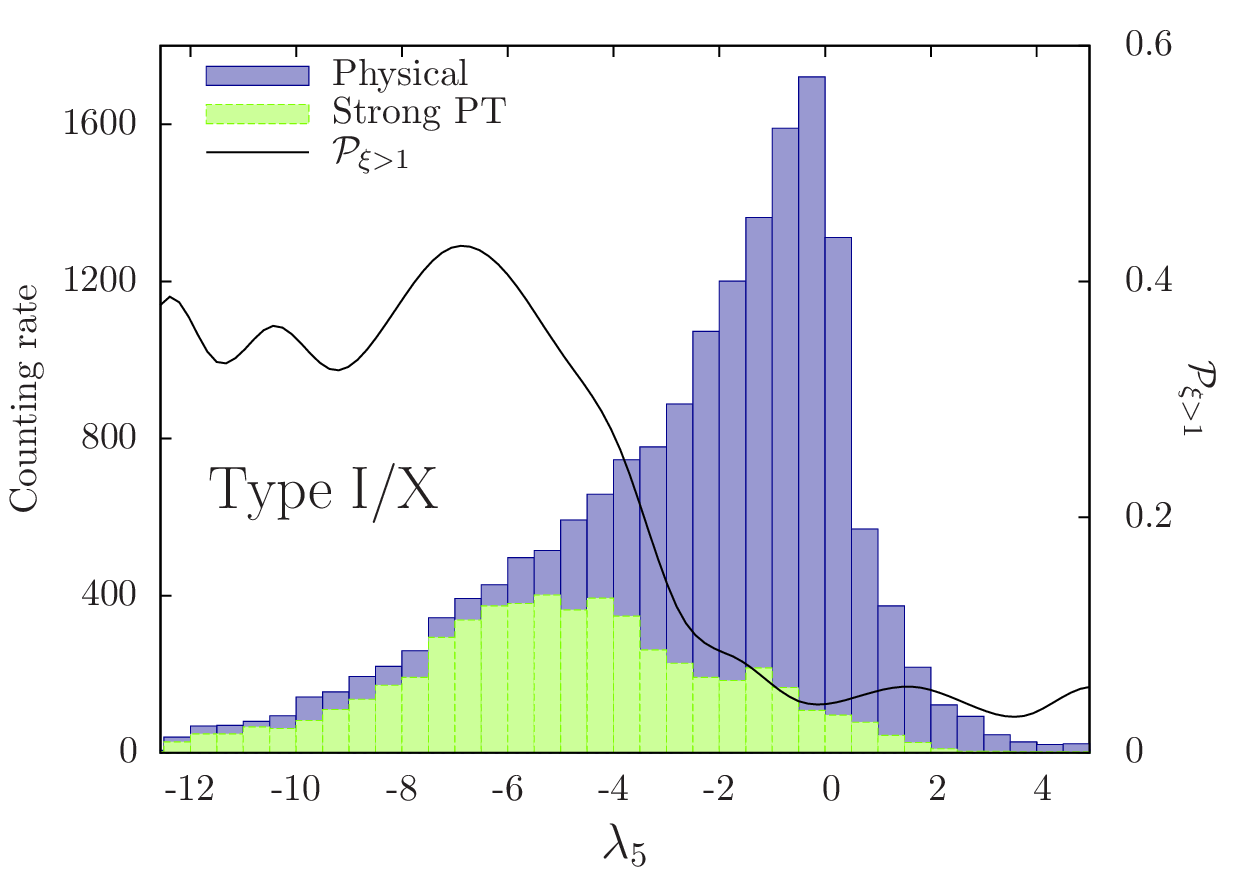}
	\includegraphics[scale=0.6]{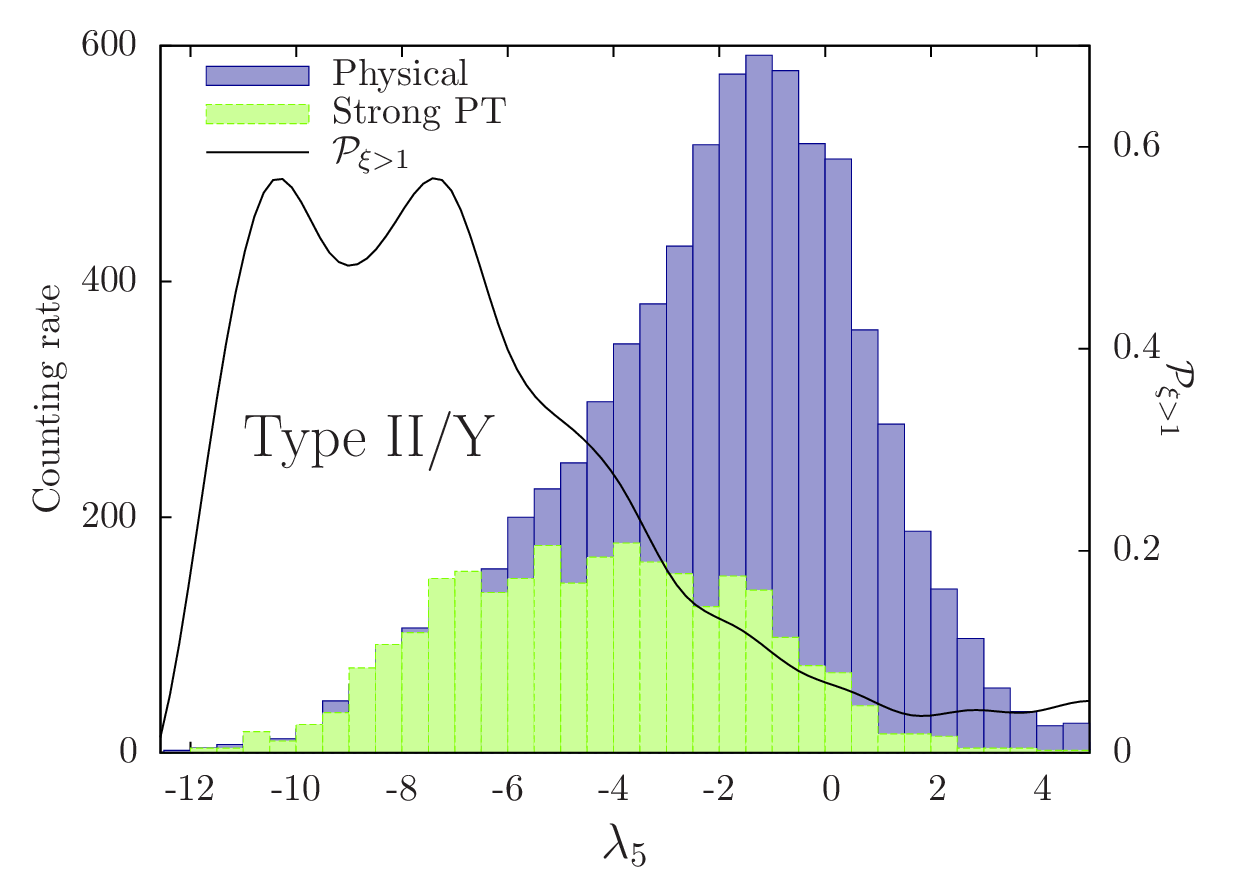}
	\caption{\small{Counting rates for physical (blue/dark) and strong phase transition (green/light) points, and their ratio, as a 
function of $\lambda_{4}$ (top) and $\lambda_5$ (bottom) in Types I/X (left) and Types II/Y (right).}}
	\label{fig:couplings}
\end{figure}

The preference for large $m_{A^0}$ indicates that the most relevant couplings for the dynamics of the 
phase transition are $\lambda_{4}$ and $\lambda_{5}$, and that they tend to be large. This is 
confirmed in Fig.~\ref{fig:couplings}. Since $\lambda_5$ regulates the splitting between $M$ and $m_{A^0}$, it 
is preferred to be big (in modulus) in both Types I/X and II/Y. As for $\lambda_4$, its general behaviour is more 
dependent on the specific model type, due to the large lower bound on $m_{H^\pm}$ that exists in Types II/Y, 
while Types I/X favour $m_{H^\pm}\approx v$. However, when the requirement of a strongly 
first-order phase transition comes into play, both types of model favour large, positive values for $\lambda_4$, 
reflecting the preferred hierarchy $m_{A^0}>m_{H^\pm}$. All these observations confirm the expectation 
that a strong phase transition requires, in general, at least some of the couplings in the model to be large. 

We see from Fig.~\ref{fig:couplings} that the couplings often approach the perturbativity limit we impose, which might cause
some concern as to the validity of our loop expansion in these regions. To be on the safe side, one could take an even smaller upper limit,
say $\lambda_i<\sqrt{4\pi}$ \footnote{A more thorough treatment of perturbativity in 2HDMs has been presented in Ref. \cite{Barbieri:2006dq}, 
showing that this upper limit may already be too conservative.}, and from Fig.~\ref{fig:couplings} 
we see that we would still have plenty of points with a strong phase transition.
Our conclusions above would also still hold, only with an upper limit of approximately $500-600$~GeV for the masses.

\subsubsection{$h^0\to\gamma\gamma$}

\begin{figure}[h!]
	\centering
	\includegraphics[scale=0.6]{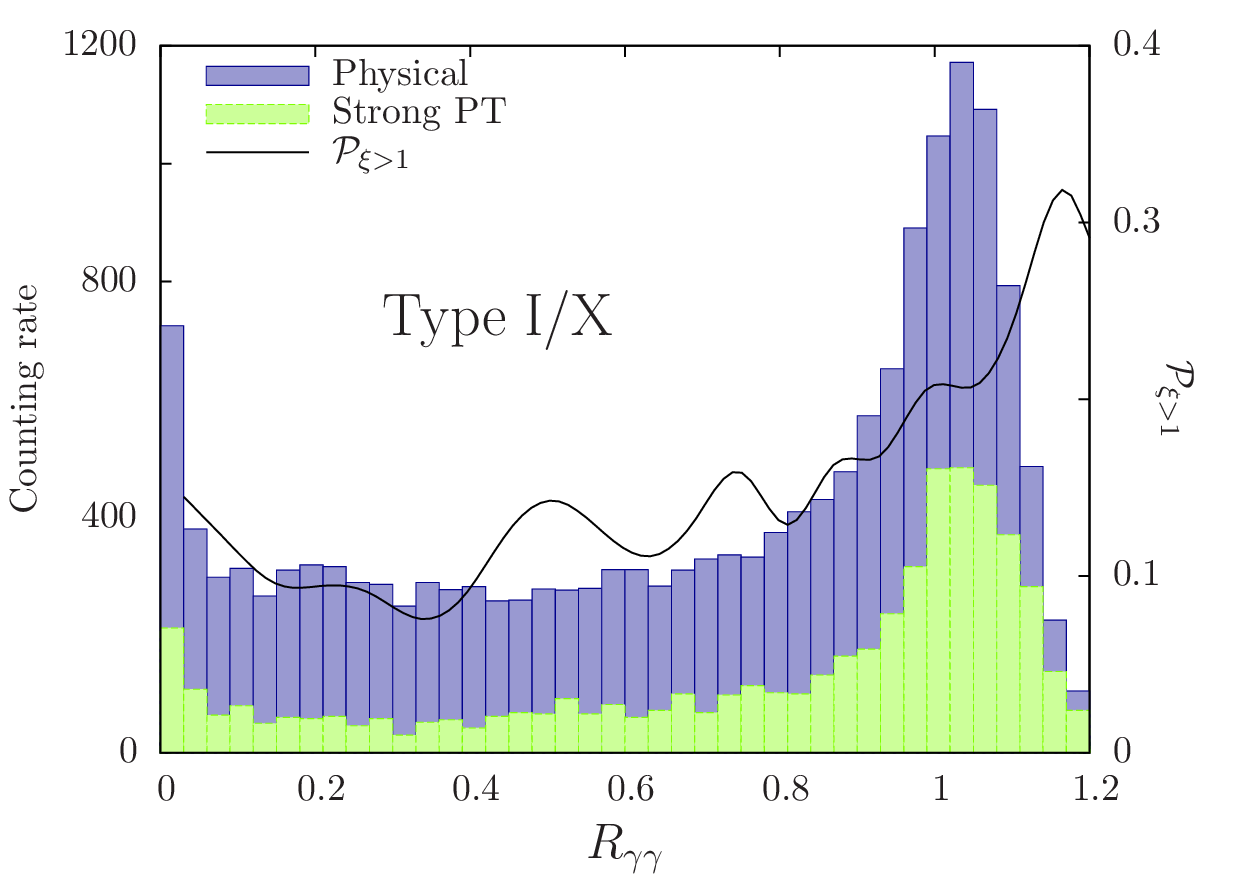}
	\includegraphics[scale=0.6]{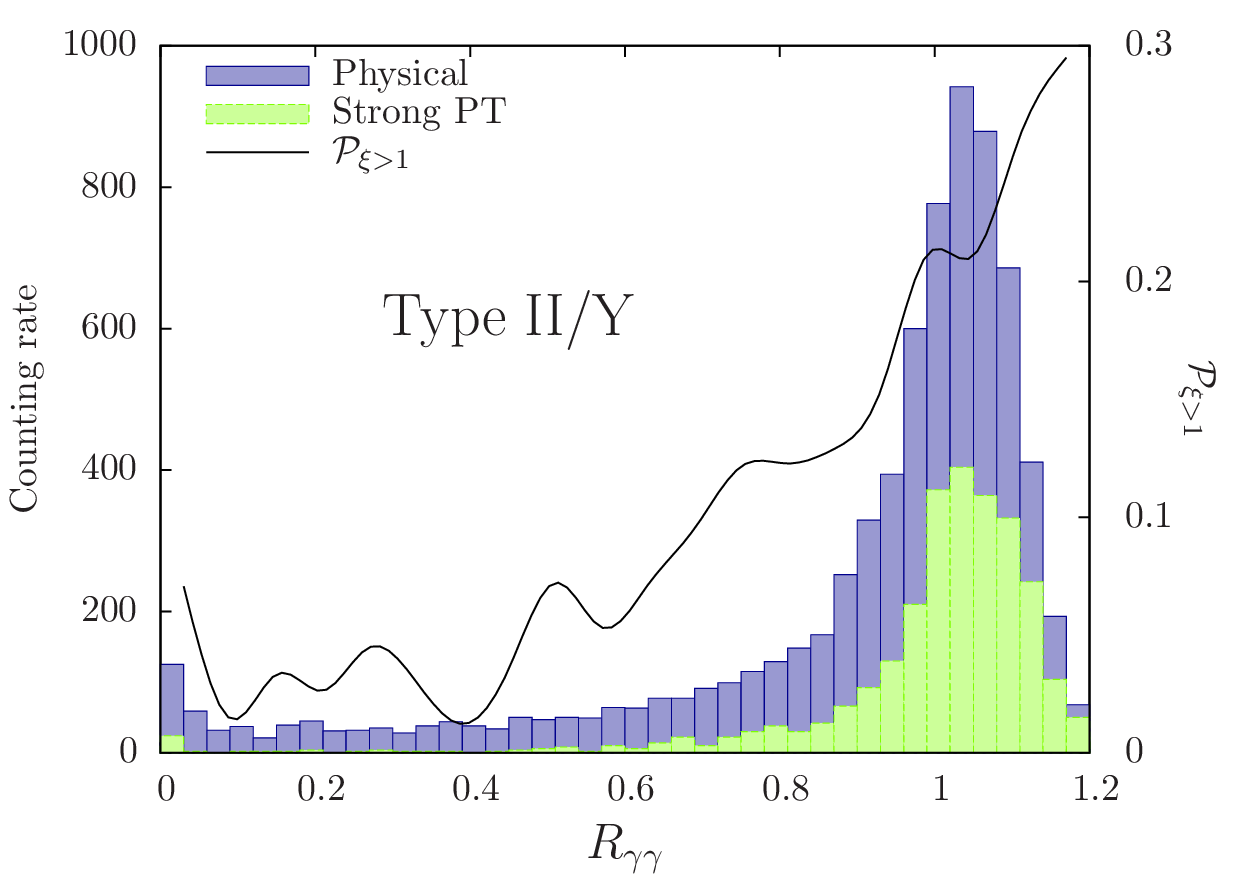}
	\caption{\small{Number of physical (blue/dark) and strong phase transition (green/light) points as a function 
of the $h^0\rightarrow \gamma \gamma$ decay width (normalized to its SM value) for Types~I/X (left) and Types~II/Y~(right).}}
	\label{fig:digamma}
\end{figure}

In 2HDMs the $h^0\to\gamma\gamma$ decay can deviate from its SM counterpart due to a difference in 
the $h^0$ couplings to $W^\pm$ and fermions, as well as to the existence of an extra charged particle 
mediating the process, namely $H^\pm$. The latter can either enhance or suppress the contribution of the 
former, and we find that this behaviour is determined mainly by $\beta-\alpha$, with an enhancement 
favoured by $\alpha\approx\beta$. 
Fig.~\ref{fig:digamma} shows the decay width of $h^{0}\rightarrow \gamma \gamma$ normalized to 
the SM rate~\cite{Djouadi:2005gj, Posch:2010hx},
\[ R_{\gamma\gamma}\equiv \frac{\Gamma_{2HDM}(h^0\to\gamma\gamma)}{\Gamma_{SM}(h^0\to\gamma\gamma)}.\]
As expected, there is a peak around the unit value in the counting rate distributions due to the preference for a 
SM-like $h^{0}$, i.e. $\alpha\approx \beta$. This is even more pronounced in Types~II/Y (cf. 
Fig.~\ref{fig:b-a}). Furthermore, a strong phase transition scenario 
favours $\alpha\approx\beta$ even more sharply, thus favouring an enhancement in the $h^0\to\gamma\gamma$ width 
from $H^\pm$ contributions in the loop, as seen in the lines for $\mathcal{P}_{\xi>1}$. 

\begin{figure}[h!]
	\centering
	\includegraphics[scale=0.6]{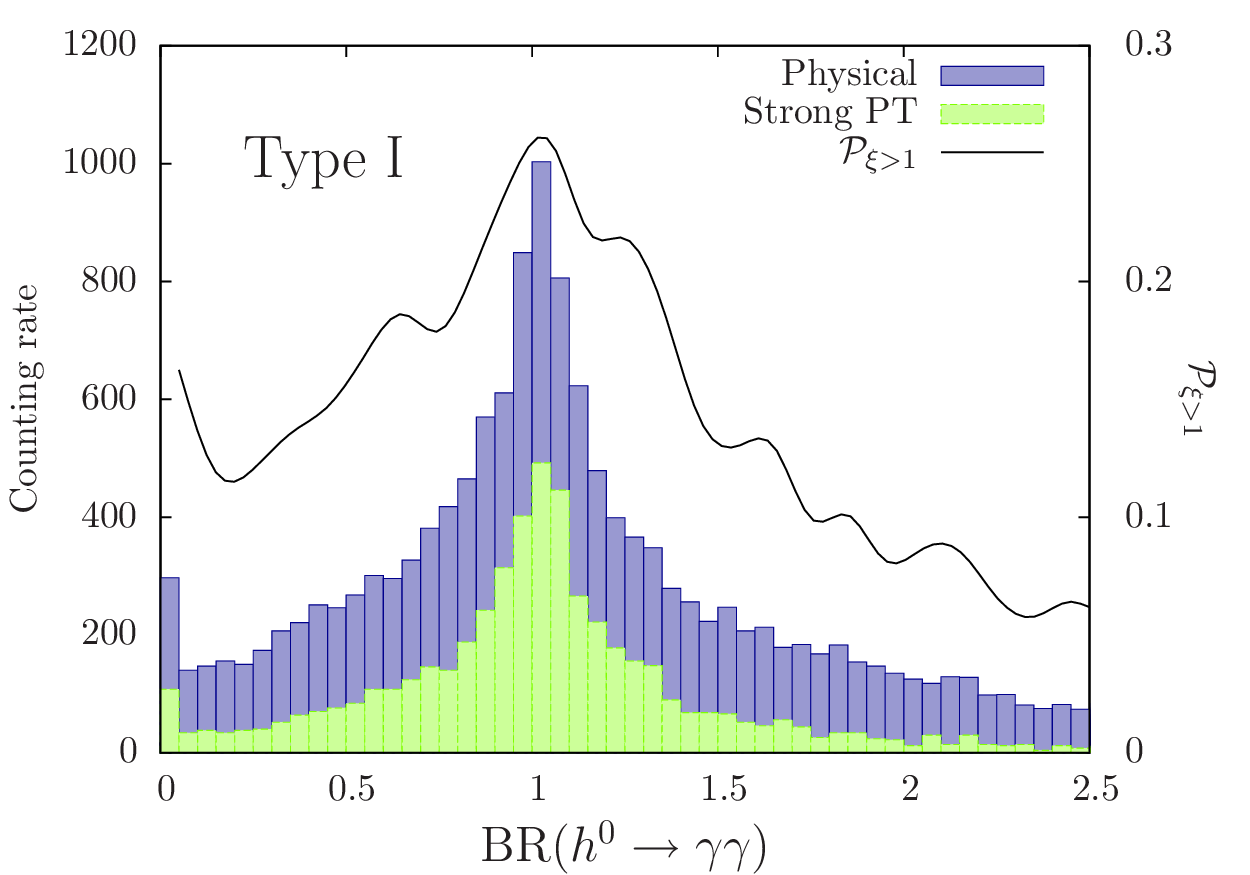}
	\includegraphics[scale=0.6]{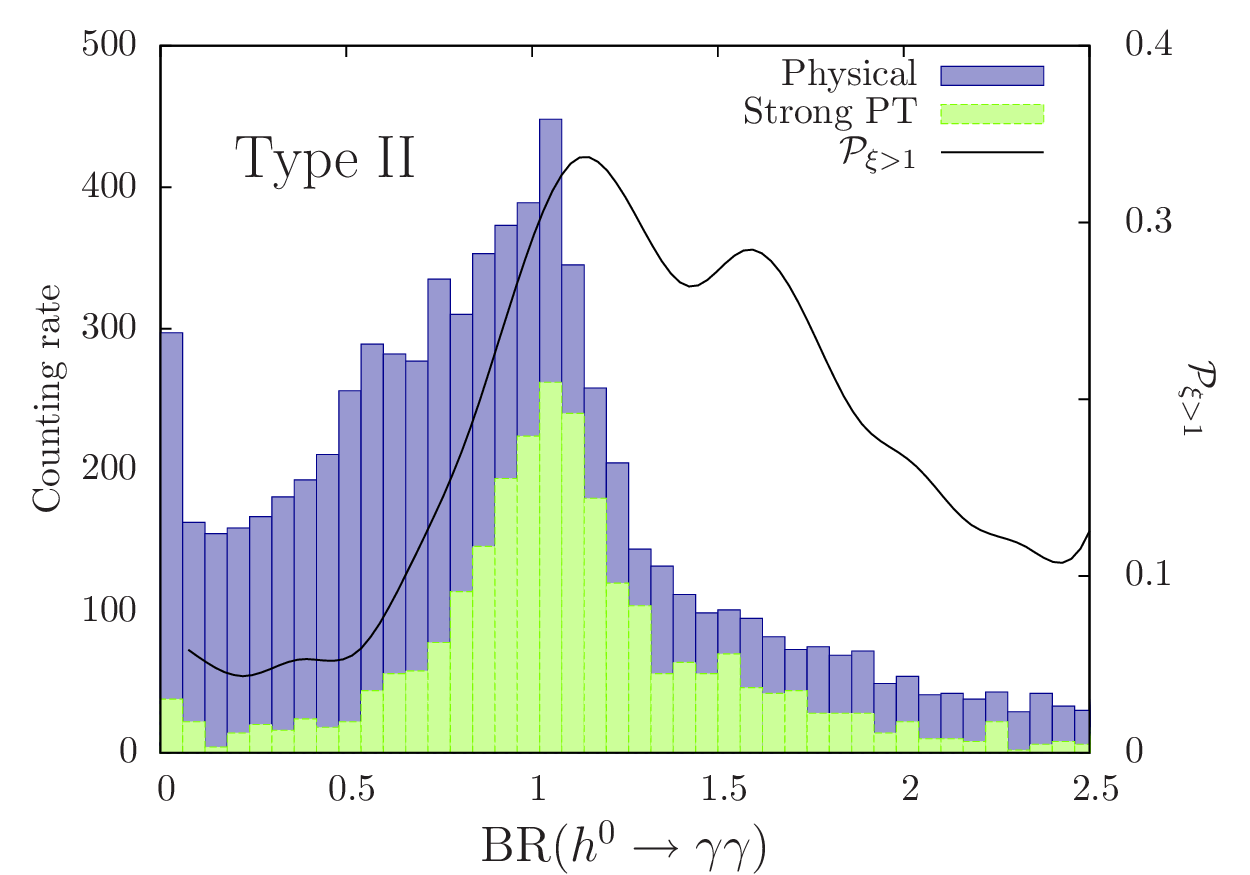}
	\caption{\small{Number of physical (blue/dark) and strong phase transition (green/light) points as a 
function of the branching ratios for $h^0\to\gamma\gamma$ (normalized to their SM values) for 
Types~I/X (left) and Types~II/X (right).}}
	\label{fig:BR}
\end{figure}

In Fig.~\ref{fig:BR} we show the behaviour of the branching ratio for this decay channel in Types~I and 
II (normalized to their SM values). For the total decay width of $h^0$ we 
consider the $b\bar{b},~\tau\tau,~gg,~WW,~ZZ$ and $\gamma\gamma$ channels, whose widths are computed from 
their SM values, taking into account only the leading-order corrections due to the change 
in their couplings to $h^0$, except in the $\gamma\gamma$ case, where we consider the contribution 
from the additional particles propagating in the loop, as just discussed. The dependence on the specific model type 
enters in $\Gamma(h^0\to b\bar{b})$ and $\Gamma(h^0\to \tau\tau)$, which in Types I and II get corrected by a 
factor of\\
\[ \text{Type I: }\frac{\sin\alpha}{\sin\beta}~,\hspace{5mm}
    \text{Type II: }\frac{\cos\alpha}{\cos\beta}~. \]
Also note that, because of the $\tau\tau$ channel, Types~I and X and Types~II and Y can no longer be treated 
as indistinguishable. However, because the SM branching ratio to $b\bar{b}$ largely dominates over 
the $\tau\tau$ one, the conclusions drawn here essentially hold for Types X and Y as well. 

We have previously seen that a strong phase transition favours $\alpha\lesssim \beta$. This means 
that in Type~I the $b\bar{b}$ and $\tau\tau$ decays are enhanced, and, as a consequence, 
the preference for an enhanced $h^0\to\gamma\gamma$ width 
is counterbalanced by an simultaneous enhancement in the Higgs' total width, so that there is an
overall preference for BR$(h^0\to\gamma\gamma)\approx1$.
The very opposite happens in Type~II, where not only an enhanced $\gamma\gamma$ width is 
favoured, but also the total width is suppressed, thus increasing the branching ratio for this channel. 
We nevertheless emphasize that in both cases a slight enhancement in the $\gamma\gamma$ channel is viable 
in a strongly first-order phase transition scenario.

\section{\label{section:conclusions}Conclusions}

Our scan over a wide range of the 2HDM's parameter space shows that these models are 
robust candidates for accomodating a strongly first-order electroweak phase transition in light of the recent 
LHC results. In particular, a strong phase transition scenario is favoured if the lightest scalar of the 
model behaves like the SM Higgs, as is indeed the case of the recently observed resonance. 
Put another way, the discovery of such SM-like scalar constrains the parameter space of 2HDMs to regions 
that favour strongly first-order phase transitions. 

The main results from our analysis are that a strongly first-order phase transition prefers: 
(a) $\tan\beta\approx 1$, which is excellent from the baryogenesis perspective, as the baryon 
asymmetry generated in the model is suppressed 
with $n_B\sim(\tan\beta)^{-2}$; (b) $\alpha\approx\beta$ (with a slight tendency for $\alpha<\beta$), in very good agreement with
the results from the $7$ and $8$ TeV runs of LHC; (c) a mass hierarchy in the scalar 
sector, $m_{H^\pm}\lesssim m_{H^0} < m_{A^0}$; (d) a rather heavy pseudo-scalar, $m_{A^0}\gtrsim 400$ GeV, this being in fact the most influential parameter on the dynamics of the phase transition.

We also find that, in Type II (and also Type Y) models with a strongly first-order phase transition, 
an enhancement in the digamma branching ratio of the lightest neutral scalar is preferred. 
In Type I (and Type X) models the opposite occurs, and the tendency is for a suppression of this branching ratio.

It is important to point out that apart from the requirement of a strongly first-order 
electroweak phase transition, for baryogenesis the wall velocity of the expanding bubbles $v_w$
plays an important role. Using the analysis from~\cite{Huber:2013kj} 
we expect in the present case a wall velocity $v_w \sim 0.2$ for mildly strong phase 
transitions ($\xi \gtrsim 1$), as the extra scalar degrees of freedom tend to slow down the bubble 
walls compared to the SM case, which favours the electroweak baryogenesis scenario in 2HDMs. In contrast, 
for very strong phase transitions the wall velocity may be significantly larger (but even in this case electroweak 
baryogenesis may be possible, see~\cite{Caprini:2011uz}). 
 
Finally, it would now be interesting to examine whether, and how, our results for the strong phase 
transition can help bring 2HDMs closer to being tested in the next run of the LHC. First steps 
in this direction have already been taken~\cite{Craig:2013hca,Djouadi:2013qya}.

\begin{center}
\textbf{Acknowledgements} 
\end{center}

We thank M. Ramsey-Musolf, T.~Konstandin, A. Djouadi and M.~Wiebusch for useful 
discussions. S.H. and J.M.N. are supported by the Science Technology and Facilities Council (STFC) 
under grant No.\ ST/J000477/1.

\end{document}